\documentclass[twocolumn, times, tighten]{aastex63}
\usepackage{graphicx,multirow,hyperref,url,color}
\hypersetup{colorlinks=true, urlcolor=blue, citecolor=blue}

\newcommand{\mbh}{M_{\rm BH}}
\newcommand{\msun}{{\rm M}_{\sun}}
\newcommand{\ledd}{L_{{\rm Edd}}}
\newcommand{\mdot}{\dot M}
\newcommand{\medd}{\dot M_{\rm Edd}}
\newcommand{\ergs}{\rm \,erg\,s^{-1}}
\newcommand{\rg}{{R_{\rm g}}}
\newcommand{\rmag}{{R_{\rm m}}}
\newcommand{\vvect}{{\bf v}}
\newcommand{\bvect}{\textbf{\textit{B}}}
\newcommand{\bz}{B_z}
\newcommand{\br}{{B_R^{\rm s}}}
\newcommand{\bphi}{{B_{\phi}^{\rm s}}}
\newcommand{\pr}{\mathcal{P}_{\rm m}}
\newcommand{\betazrout}{{\bar{\beta}_{z0}}}

\begin{document}
\defcitealias{Narayan2003}{N03}
\defcitealias{Igu2008}{I08}
\defcitealias{Narayan2012}{N12}
\defcitealias{McKinney2012}{M12}

\title{Radiative Properties of Magnetically-Arrested Disks}
\shorttitle{Magnetically-Arrested Disks}

\author[0000-0001-9969-2091]{Fu-Guo Xie}
\affil{Key Laboratory for Research in Galaxies and Cosmology, Shanghai Astronomical Observatory, Chinese Academy of Sciences, \\
80 Nandan Road, Shanghai 200030, China; \href{mailto:fgxie@shao.ac.cn}{fgxie@shao.ac.cn}}

\author{Andrzej A. Zdziarski}
\affil{Nicolaus Copernicus Astronomical Center, Polish Academy of Sciences, Bartycka 18, PL-00-716 Warszawa, Poland; \href{mailto:aaz@camk.edu.pl}{aaz@camk.edu.pl}}

\shortauthors{Xie \& Zdziarski}

\begin{abstract}
Magnetically-arrested disks (MADs) appear when accretion flows are supplied with a sufficient amount of magnetic flux. In this work, we use results of magnetohydrodynamic simulations to set the configuration of the magnetic field and investigate the dynamics and radiative properties of the resulting accretion flow (i.e., without that of the jet) of MAD. The method developed here is applied to both the MAD and the standard and normal evolution (SANE) accretion flow with or without large scale magnetic fields. For the radiative processes, we include synchrotron, bremsstrahlung and Compton scattering. We find that, in general, MAD accretion flows have similar spectra to those of the SANE, which complicates the task of distinguishing MADs from SANEs. At the same accretion rates, MADs are systematically brighter than SANEs. However, the critical accretion rate above which the hot solution ceases to exist is lower in MAD. Consequently, the maximum luminosity the MAD can reach is comparable but slightly lower than that of SANE, and the dependence on the magnetic flux is weak. We then discuss implications of our results for active galactic nuclei and accreting black-hole binaries.
\end{abstract}


\section{Introduction}
\label{sec:intro}

Active galactic nuclei (AGNs) and black hole (BH) X-ray binaries (BHBs) are believed to be powered by the accretion of gas onto the central BH, a process that can efficiently convert gravitational potential energy and BH spin energy into radiation. According to the gas temperature, accretion disks can be crudely classified into two categories \citep{Yuan2014}, one is a geometrically-thin and optically-thick cold disk emitting blackbody radiation, first developed by \citet[][herefter abbreviated as SSD]{SS1973}, and the other is a geometrically-thick hot accretion flow with low optical depth and thus with optically-thin emission (though still by predominantly thermal electrons), e.g., an advection-dominated accretion flow \citep[ADAF;][]{Narayan1994}. Accretion disks usually contain small-scale turbulent magnetic fields (with length scales much shorter than the disk height, $H$), which are generated and amplified by the magnetohydrodynamic (MHD) turbulence such as the magneto-rotational instability (MRI; \citealt*{Balbus1998}). The MRI-driven turbulent field is usually weak, with the gas-to-magnetic pressure ratio, $\beta$, parameter being greater than unity. Although crucial for the angular momentum transfer \citep{Balbus1998}, MRI has a weak impact on other dynamical properties of the accretion disk.

In above standard picture, which is also called the ``standard and normal evolution'' (SANE; \citealt{Narayan2012}, hereafter \citetalias{Narayan2012}; \citealt*{Sadowski2013}) for sub-Eddington hot accretion case, the existence of global large-scale magnetic fields (with the length scale longer than $H$) around the disk is mostly neglected\footnote{For magnetic fields, we hereafter use the terms `turbulent' and `small-scale', and `global' and `large-scale' interchangeably.}. This is because it is difficult for the MRI or dynamo processes to generate stable (in the sense of time sustained rather than temporary) global magnetic field. Stable global magnetic fields can only be advected/dragged inward from the environment by accreting material (for the pioneering work, see \citealt*{B1974}, and \citealt*{B2019} for a recent review)\footnote{We note that alternative mechanisms are proposed in literature to generate a significant amount of magnetic flux locally within the accretion disk, e.g., an accumulating random magnetic flux supply by SSD at its inner truncation radius, $R_{\rm tr}$, inside of which the disk becomes hot \citep{Begelman2014}.}, but they will diffuse outward, and the advection-diffusion competition depends on the turbulent magnetic Prandtl number (the ratio of the viscosity to the magnetic resistivity) \citep{Guan2009}. The field accumulation is found to be sensitive to the disk aspect ratio $H/R$ (where $R$ is disk radius); namely, it is efficient in hot thick flows such as the ADAF, but will be highly suppressed in cold thin disks like SSD (e.g., \citealt{Guilet2012}). 

Only until the 2000s did the community recognize the importance of the supply and accumulation of magnetic flux in the accretion disks. The magnetic flux, $\Phi$, is defined as a surface integral of magnetic field vector $\textbf{\textit{B}}$, i.e. $\Phi=\int \textbf{\textit{B}}\cdot {\rm d}\textbf{\textit{S}}$. There are three main reasons for this change. The first is that compared to the turbulent magnetic field, the global one has the advantage that it cannot be accreted into the BH, nor can it be dissipated locally due to magnetic diffusivity (\citealt*{Igu2008}, hereafter \citetalias{Igu2008}). Once supplied, the global field can be stored and accumulated for a long time, e.g., the duration of an occurrence of the hard state in a BHB. Even if the accretion disk is supplied by fluxes with random signs, strong global magnetic field can develop (e.g., \citealt{Narayan2003}, hereafter \citetalias{Narayan2003}; \citealt{Begelman2014}). The second is the existence of significant coherent magnetic flux in the interstellar medium near AGNs, which makes flux feeding scenario plausible. For AGNs, the transverse Faraday rotation gradient in the jets provides a direct signature of magnetic polarity (e.g., \citealt*{Gabuzda2015}). The core-shift effect detected from high-angular-resolution multi-band radio observations also provides estimates of the magnetic flux (e.g., \citealt*{Z2014, Zdziarski2015}). Rough estimatest of the mean strength and coherent length of magnetic fields in the interstellar medium also favor a magnetic flux as large as $\Phi \sim 0.1 {\rm pc^2\,G}$ \citepalias{Narayan2003}. For BHBs, the field strength at the surface layer of the donor star can be as high as $B\sim$100--1000 G \citep{Reiners2009, Reiners2010}. Accretion of material from the donor, either through stellar wind or through Roche-lobe overflow, may possibly supply a flux $\Phi\sim 10^{-14}$--$10^{-13}\, {\rm pc^2\,G}$ \citep{Sadowski2016}.

The third reason, related to the focus of this work, is the discovery of the ``magnetically-arrested disk'' (MAD; \citealt*{Igu2003}; \citetalias{Narayan2003}). It is also called the ``magnetically choked accretion flow'' (\citealt*{McKinney2012}, hereafter \citetalias{McKinney2012}). All magnetized accretion flows will accumulate a certain amount of magnetic flux near BH. However, if the disk is fed continuously with a sufficiently large amount of magnetic flux, it will evolve into a MAD phase, where the magnetic force is strong enough to balance the inflow's ram or gravitational force at certain location. This location is defined as the magnetospheric radius $\rmag$. Taking $G\mbh\Sigma/R^2 \sim B_R B_z/4\pi$ (where $\mbh$ is the BH mass and $\Sigma$ is the disk half-column density; see Equation\ \ref{eq:gm}), $\rmag$ can then be estimated as (Sec.\ \ref{sec:rmag}; see \citetalias{Narayan2003} for a simplified expression without considering the effect of $H/R$),
\begin{eqnarray}
  {\rmag\over \rg} & \approx & \pi^{-4/3} \alpha^{2/3} \left({H\over R}\right)^2 \dot{M}^{-2/3} \left(R_{\rm g}^2 c\right)^{-2/3} \Phi^{4/3}, \label{R_m} \\
  & \approx & 5.3\times10^4 \alpha_{-1}^{\frac{2}{3}} \dot{m}_{-1}^{-\frac{2}{3}} m_8^{-2} \left({H/R\over0.5}\right)^2 \left({\Phi\over 0.1 {\rm pc^2\,G}}\right)^{\frac{4}{3}},\nonumber \\
  & \approx & 5.3\times 10^2 \alpha_{-1}^{\frac{2}{3}} \dot{m}_{-1}^{-\frac{2}{3}} m_1^{-2} \left({H/R\over0.5}\right)^2 \left({\Phi\over 10^{-13} {\rm pc^2\,G}}\right)^{\frac{4}{3}}.\nonumber
\end{eqnarray}
Here, $m\equiv \mbh/\msun$, $\dot{M}$ is the mass accretion rate and $\dot{m}$ is the accretion rate in Eddington unit, $\dot{m} = \dot{M}/\medd$\footnote{We define $\medd=10\ledd/c^2$ as the Eddington accretion rate, where $\ledd\approx 1.3 \times10^{46}\ergs (\mbh/10^8 \msun)$ is the Eddington luminosity for pure H.}, a quantity $X_n$ is defined as $X/10^n$, $\alpha$ is the dimensionless viscosity parameter \citep{SS1973} and $\rg = G\mbh/c^2$ is the gravitational radius. For the expressions above, we also take $\Phi \approx \pi R_{\rm m}^2 \bz$, $\Sigma = \rho H = -\dot{M}/4\pi R V_R$, where the radial velocity $V_R$ is estimated as $V_R \approx -\alpha (H/R)^2 R\Omega_{\rm K}$ \citep{Narayan1994}, where $\Omega_{\rm K}$ is the Keplerian angular velocity, and we further assume $B_R/B_z\sim H/R$ (see Equation (\ref{eq:br}) below and \citealt*{Lubow1994}). We note that the expression above neglects possible outflows; if they are present, $\dot M$ above would be that at $R_{\rm m}$. For hot accretion flows, SANEs are then generalized to include systems whose magnetic fluxes remain below a critical value \citepalias{Narayan2012}, which can be derived by setting $\rmag$ to the BH horizon radius. 

\setlength{\tabcolsep}{3pt}
\begin{table}
\centering
\caption{Model Definitions/Abbreviations}
\begin{tabular}{cl}
\hline
\hline
Abbreviation & Definition\\
\hline
SANE & Standard And Normal Evolution (of hot flow) \\
standard SANE & SANE with extremely weak global fields\\
near-critical SANE & SANE whose $\rmag\lesssim2\rg$ (BH horizon)\\
\hline
MAD & Magnetically-Arrested (hot) Disk \\
low-$\Phi_{\rm n}$ MAD & MAD which has $2\rg<\rmag\la (10-20)\rg$ \\
high-$\Phi_{\rm n}$ MAD & MAD whose $\rmag \ga R_{\rm out}$\\
\hline
\end{tabular}\label{tab:name}
\end{table}

We note that the MAD state can also be achieved in cold accretion disks \citep{Avara2016, Marshall2018} and in super-Eddington disks \citep{McKinney2015}. In this work, we limit ourselves to the sub-Eddington hot ADAF-like disks, namely the ADAF/MAD. For simplicity, below we name ADAF/MAD as MAD, and ADAF/SANE as SANE. As listed in Table \ref{tab:name}, we additionally define two extreme cases, one is the ``low-$\Phi_{\rm n}$ MAD'' (where $\Phi_{\rm n}$ is a normalized magnetic flux, see Sec.\ \ref{sec:mag} below) with a small value of $\rmag$ ($2\rg<\rmag\la (10$--$20)\rg$), and the other is the ``high-$\Phi_{\rm n}$ MAD'', whose magnetic flux is so large that $\rmag$ is greater than the outer boundary of the hot accretion disk, $R_{\rm out}$. For comparison, we additionally specify the SANE with negligible global fields as the ``standard SANE'', and the SANE with maximal magnetic flux (before transit to the MAD state, namely $\rmag \lesssim 2\rg$) as ``near-critical SANE''. 

MAD has several notable advantages compared to SANE. First, magnetic field topology is crucial for the jet production (e.g., \citealt*{Igu2003, Beckwith2008, McKinney2009}), and MAD is the ideal site in extracting the BH spin energy and launching the Blandford-Znajek jet (BZ-jet; \citealt*{BZ77}). The jet in MAD is systematically more powerful than that produced in SANE (\citetalias{Narayan2012}; \citealt*{Sadowski2013}). The jet efficiency, defined as the ratio of the jet power to the accretion power, can exceed $100\%$ in MADs (e.g., \citealt*{Tch2011}), while the jet efficiency of SANE is usually less than $20\%$ (e.g., \citealt*{Beckwith2008, Sadowski2013}). Statistical analysis of radio-loud AGNs suggests that they are in the MAD state \citep{Z2014, Ghisellini2014, Zdziarski2015}. Secondly, MAD systematically has a higher radiative efficiency compared to SANE under the same $\dot{M}/\medd$, because of higher surface density (see Sec. \ref{sec:lcrit}). Consequently, a brightening of a system can be caused by the feeding and accumulation of magnetic flux, while it is would usually be attributed to an increase in the mass accretion rate. Thirdly, MAD may explain the timing properties in BHBs. For example, from 3D MHD simulations, \citetalias{McKinney2012} proposed that high-frequency quasi-periodic oscillations (QPOs) observed in BHBs can be triggered in MAD systems due to the magnetic interchange and Kelvin--Helmholtz instabilities at the disk-jet interface, and \citet{Tch2011} proposed that low-frequency QPOs in BHBs occur due to fluctuations of the radiative efficiency of MADs\footnote{Alternatively, the low-frequency QPOs may be driven by other mechanisms such as Lense-Thirring precession due to spin-disk misalignment, see e.g., \citet*{Ingram2009, Liska2019}.}. 

One obstacle that hinders the determination whether an observed source is a MAD is the lack of spectral calculation. Most MAD simulations in literature focus on the accretion dynamics and/or the jet properties, without considering the radiation part. Such simplification limits their application to bright systems whose accretion rates (in Eddington unit) are high. It is known from simulations of SANEs that radiative cooling plays an important role, i.e. the density and temperature of the flow will be significantly changed (between radiative and non-radiative cases) at a rather low $\dot{M}/\medd$ \citep{Dibi2012}, and the wind/outflows are highly suppressed at higher $\dot{M}/\medd$, at which the radiative cooling is comparable to the viscous heating \citep{Wu2016, Bu2018}. Besides, the coupling between electrons and ions is weak in hot accretion flows. Consequently, the electrons, which produce most of the radiation, have a temperature significantly different to that of ions \citep{Narayan1995}, i.e. the accretion is two-temperature. Accretion flows with both global fields and two-temperature plasma have been investigated for SANEs (e.g., \citealt*{Oda2007, Oda2012, LC2009, Cao2011, Wu2016, Sadowski2017, Bu2018}; though only few provide spectra, e.g., \citealt*{Oda2012, Ryan2018}) but rarely for MADs. The only exception to our knowledge is \citet*{Chael2019}, who carried out two-temperature, radiative MHD simulations of MAD systems and further calculated the spectrum. In their work, they focused on low $\mdot/\medd$ cases, targeting M87.

In this work, we investigate radiative properties of MAD and compare them with those of SANEs. Section \ref{sec:madmodel} is devoted to the basic properties of MAD learned from previous simulations, together with a presentation of our model setup. We assume the MAD is stationary, and use the height-integrated equations. With these simplifications, we explore the impact of various adopted parameters. Section \ref{sec:result} then presents detailed numerical calculations, Section \ref{sec:discussion} is devoted to discussions of MAD state in various systems, while Section \ref{sec:summary} presents a brief summary.

\section{Basic Properties Of MAD And Model Setup} \label{sec:madmodel}

\subsection{MAD Dynamics in MHD Simulations} \label{sec:madsim}

Same with SANE, MAD also consists of a magnetized polar relativistic jet and an equatorial inflow with outflowing gas above its surface. Below we summarize the basic properties of MAD, based on state-of-art 3D MHD simulations in literature (\citetalias{Igu2008, Narayan2012, McKinney2012}; \citealt*{McKinney2015, White2019}). Since the jet physics (acceleration, mass loading, proton/lepton composition, energy dissipation, etc.) is far from mature (for reviews of jet dynamics, see \citealt*{Hawley2015, Tch2015}), and the existing jet models are rather phenomenological, in this work, we study the accretion flow only.

\begin{figure*}
\centering
\includegraphics[height=6.cm]{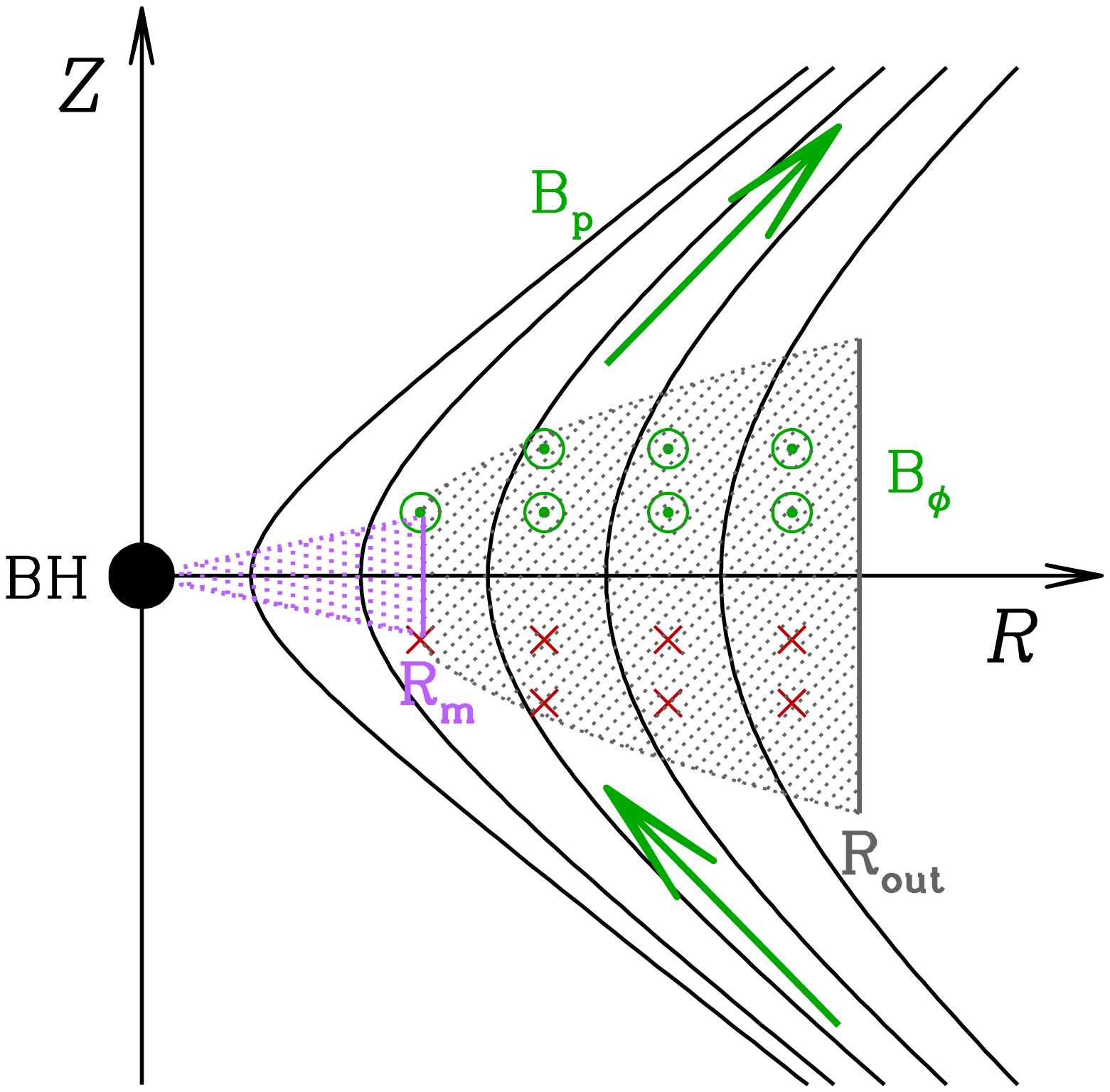}\hspace{2cm}
\includegraphics[height=6.cm]{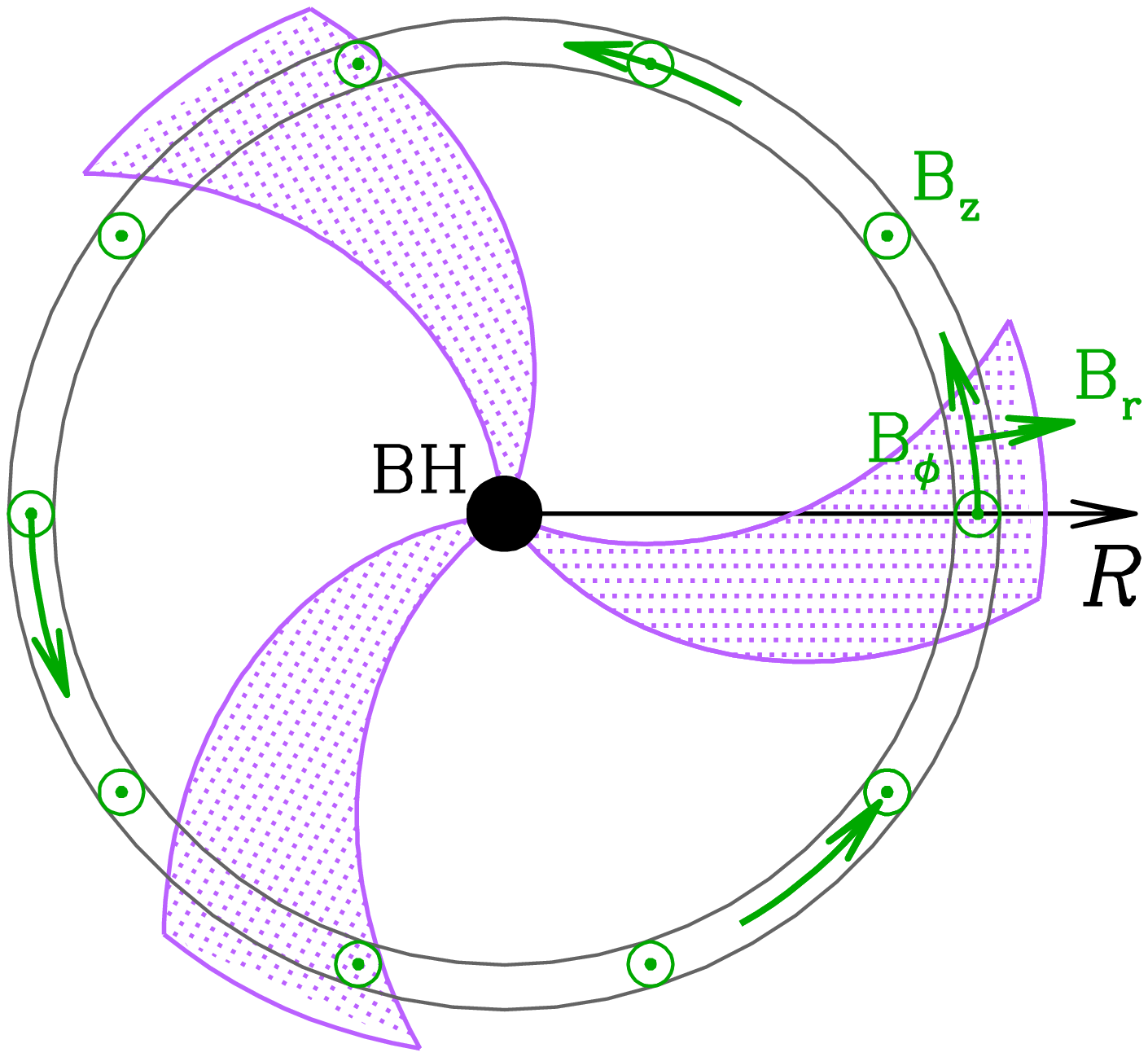}
\caption{Schematic plot of the MAD structure (mass and magnetic fields) in both the $Rz$ ({\it Left Panel}; edge-on view) and the $R\phi$ ({\it Right Panel}, face-on view. Here only the inner $R<\rmag$ region is shown) planes. In both panels, the shadowed regions mark the main body of the accretion flow, among which the violet is for the nonaxisymmetric $R<\rmag$ regions. For illustrative purpose, we shown in the right panel three dense gas spirals, outside of which are the magnetized voids. For the magnetic field configuration in the {\it Left Panel}, the black solid curves show the poloidal $B_p$ (consist of both $B_R$ and $B_z$), while the green symbols ($\times$ and $\odot$ for pointing into and out of the paper) mark the direction of the toroidal $B_\phi$. As shown in this plot, $B_R$ and $B_\phi$ are odd functions of $z$, while $B_z$ is an even function of $z$. For the field in the {\it Right Panel}, the green arrows mark the radial and azimuthal magnetic components, while the $\odot$s mark the vertical one. For the guiding numerical simulations of this setup, see e.g.\ Fig.\ 1 of \citet{Tch2011} and Figs.\ 4 and 13 of \citetalias{McKinney2012}.}
\label{fig:schemeplot}
\end{figure*}

Far outside of the magnetospheric radius, at $R\gg \rmag$, MAD shares similar dynamical and radiative properties with the high-$\Phi_{\rm n}$ SANE, although MAD usually exhibits weaker turbulence \citepalias{McKinney2012}. However, there are dramatic changes near $\rmag$. The magnetic force is so strong that the trigger of magnetic interchange instability (MII) will become inevitable (e.g., \citealt*{Tch2011}, \citetalias{Narayan2012, McKinney2012}; \citealt*{White2019}). The gas will be accreted in the form of irregular dense filamentary shears/streams with a spiral shape, and hereafter we call them the ``gas spiral streams" or more briefly ``spirals''. The dense gas spirals are under-magnetized, with gas-to-magnetic pressure ratio of $\beta\ga 1$, where $\beta$ is defined including all components of the magnetic field. The global magnetic fields are split into separate bundles, and they will penetrate through the disk plane in low-density regions \citepalias{Igu2008}. Below we will call these dilute but highly magnetized regions, whose $\beta\sim 0.01$ \citepalias{Igu2008, McKinney2012}, as the ``magnetized voids'' or more briefly the ``voids'' (also named magnetic buoyant bubbles in \citealt*{White2019}). The split into the under-magnetized spirals and the magnetized voids is a direct consequence of the efficient MII \citepalias{McKinney2012}. Those MAD simulations also reveal that the dominant magnetic field component is the poloidal one. The toroidal component is suppressed, because the rotational velocity in MAD systems is much lower than that in SANEs \citepalias{Narayan2012}. Although the gas spirals and the magnetized voids are separated spatially, they are still coupled to each other. Roughly they are in a pressure equilibrium state. Moreover, at zeroth order the magnetized voids corotate with the gas spirals \citepalias{McKinney2012}. These information will be used in our model design, see Sec.\ \ref{sec:model}.

For completeness, we note that when spiraling inward, the accreting gas drags the magnetic field and twists it around the rotation axis. Such magnetic flux will eventually thread the BH instead of the accretion flow. Collimated Poynting jets in the polar directions are then launched. In the MAD system, the magnetic flux threading the BH will oscillate around a saturated (maximal) value \citep{Tch2011}. The saturation is due to the force balance mediated by nonaxisymmetric MII \citepalias{McKinney2012}. The saturation in the magnetic flux threading onto BH has been supported by observations of a sample of radio-loud AGNs \citep{Z2014}, although the scatter of individual source is rather large \citep{Zdziarski2015}.

A direct comparison of the dynamics between MADs and SANEs has been made by \citetalias{Narayan2012}. They found that for regions outside of $\sim30\rg$ the angular momentum of MAD is substantially smaller and the inflow velocity of MAD is substantially larger, compared to those of SANE. 

\subsection{Model Setup}\label{sec:model}

We aim at constructing height-integrated equations to describe the quasi-steady ($\partial/\partial t \approx 0$) optically-thin accretion flows, which can be applied to MADs and also SANEs containing global magnetic fields.\footnote{SANEs with global magnetic fields are investigated by, e.g., \citet*{Oda2007, Oda2012, LC2009, Cao2011, Ryan2017, Ryan2018}. The magnetic field setup (i.e.\ toroidal vs.\ poloidal radial distribution) in these models is different.} We adopt the cylindrical coordinates ($R, \phi, z$), and assume the pseudo-Newtonian potential \citep{PW1980} to mimic the general relativistic (GR) effect of a Schwarzschild BH, i.e. $\psi(R,z)= - G M_{\rm BH}/\left[(R^2 + z^2)^{1/2} - 2\rg\right]$.
 
For illustrative purpose, we show in Figure \ref{fig:schemeplot} the schematic configuration of MAD in both the $Rz$ ({\it Left Panel}) and the $R\phi$ ({\it Right Panel}) planes. In both panels, the shadowed regions mark the main body of the accretion flow, among which the violet highlights the $R<\rmag$ nonaxisymmetric regions. The nonaxisymmetric region is demonstrated in the {\it Right Panel} by three dense gas spirals. We include all the three magnetic field components, $B_R$, $B_\phi$, and $B_z$, and define $B$ by $B^2=B_R^2+B_\phi^2+ B_z^2$. The black solid curves in the {\it Left Panel} show the poloidal component (consisting of $B_R$ and $B_z$), while the green symbols ($\times$ and $\odot$) mark the configuration of the toroidal component, $B_\phi$. For the magnetic fields in the {\it Right Panel}, the green arrows mark the radial and azimuthal magnetic components, while the $\odot$ symbols mark the vertical component.

All the quantities are defined in the comoving frame at the midplane ($z=0$), unless explicitly stated. We further define quantities in the dense gas spirals by subscript `s', and those in the magnetized voids by subscript `v'. 

\subsubsection{Dynamical Equations for the Gas in MAD System}\label{sec:maddyn}

Below we list the dynamical equations of MAD system. The details of the derivation of these height-integrated equations can be found in {\it Appendix}\ \ref{app:madeqs}, where additional assumptions/simplifications are also provided. Here we focus on regions occupied by gas (e.g. inside of $\rmag$ only the gas spirals); the coupling between the gas spirals and the magnetized voids will be addressed later in Section \ref{sec:coupling}.

The mass accretion rate is $\mdot(R) \equiv -4\pi C_{\rm m} RH \rho V_R$, where parameter $C_{\rm m}$ represents the surface covering factor of the accreting gas (Sec. \ref{sec:spiralmodel} and Equation \ref{eq:c_m}), i.e. $C_{\rm m} = 1$ outside of $\rmag$ and $C_{\rm m} < 1$ inside of $\rmag$. Note that $\mdot(R) > 0$ since we have $V_R < 0$. Theoretically hot accretion flow associates with not only collimated jet but also strong sub-relativistic un-collimated wind/outflow (for SANEs, see \citealt*{Yuan2012b, Yuan2015} and references therein; for MADs, see e.g. \citetalias{McKinney2012}; Yang et al. in preparation). These winds are difficult to detect, since they are nearly fully ionized when launched. 
%Observational evidence for their existence accumulates gradually over the past decade, in both BHBs (during their hard state, e.g., \citealt*{MD2016, MD2018, Homan2016}) and LLAGNs (e.g., \citealt*{Tombesi2010, Tombesi2014, Crenshaw2012}). 
Following \citet{BB99}, we include outflow/wind and write the continuity equation as 
\begin{equation}
\mdot(R)= \mdot_0 \left({R\over R_{\rm out}}\right)^s = \mdot_{\rm BH} \left({R\over 2 \rg}\right)^s. \label{eq:mdot}
\end{equation}
Here $\mdot_0$ and $\mdot_{\rm BH}$ are the accretion rates at the outer boundary $R_{\rm out}$ and the BH event horizon, respectively. The impact of wind is determined by the outflow parameter $s$. In nature $s$ may vary with radius $R$. MHD numerical simulations of hot accretion flows show that wind is fairly strong outside of $\sim\! 20 \rg$ but is highly suppressed within this radius (e.g., \citealt*{Yuan2012a}). Detailed modeling of the X-ray emission in Sgr A* also confirms the suppression of wind near BH \citep{Ma2019}. Besides, the outflow strength for regions inside $\rmag$ may be different to that outside of $\rmag$. In this work we omit all these complexities but take $s$ as a constant, independent of $R$.

With the assumption that the MAD is in a stationary state in the vertical direction, the scale-height of the accretion disk, $H$, can be expressed\footnote{The additional correction to $H$ due to global fields, omitted here for simplicity, is considered by, e.g.\ \citet*{Cao2011}.} as $H=c_{\rm s}/\Omega_{\rm K}$. Here $c_{\rm s} = (P_{\rm tot}/\rho)^{1/2}$ is the isothermal sound speed, and the total pressure $P_{\rm tot}$ includes three contributions, namely of the gas, $P_{\rm g}$, of the turbulent field, $P_{\rm tm}$, and of the global field, $P_{\rm m} = B^2/8\pi$. The radiation pressure is negligible in hot accretion flows. We thus write $P_{\rm tot} = P_{\rm g} + P_{\rm tm} + P_{\rm m}$. Correspondingly, we introduce three plasma gas-to-magnetic parameters, one for the turbulent field component, $\beta_{\rm t}= P_{\rm g}/P_{\rm tm}$, one for the global field component, $\beta_{\rm m} = P_{\rm g}/P_{\rm m}$, and one for the total (turbulent+global) field component (this is the one mentioned in above sections), $\beta = P_{\rm g}/(P_{\rm tm}+P_{\rm m})$. Obviously, the three plasma parameters relate to each other, i.e., $1/\beta = 1/\beta_{\rm t} + 1/\beta_{\rm m}$.

MRI is suppressed (or only marginally activated) in MAD systems (\citetalias{McKinney2012}; \citealt*{White2019}). However, turbulence still develops, since on small scales MII can constantly form and disrupt MII streams/bubbles \citep{Marshall2018, White2019}. Such MII-driven turbulence is usually weaker compared to the MRI-driven one in SANEs (\citetalias{McKinney2012}; \citealt*{White2019}). In this work, we assume $\beta_{\rm t} = 10$, a typical value for SANEs with weak fields (e.g., \citealt*{Stone2001, Hawley2002}). As long as $\beta_{\rm t} \gg 1$, the exact value of $\beta_{\rm t}$ is not important, for neither the dynamics nor the radiative properties of MAD. Despite the change in the origin of the MHD turbulence (i.e. MRI driven in SANEs and MII driven in MADs), we estimate the torque due to the MHD turbulence, $\tau_{\phi R}$, through the conventional $\alpha$ description \citep{SS1973}. With the contribution from global fields excluded, we write the torque as $\tau_{\phi R} = -\alpha (P_{\rm g} + P_{\rm tm}$). 

The momentum equations in radial and azimuthal directions can be expressed as
\begin{eqnarray}
  &V_R {{\rm d} V_R\over {\rm d}R} + {1\over\rho} {{\rm d} P_{\rm tot}\over {\rm d}R} + {1\over 2}{{\rm d} c_{\rm s}^2\over {\rm d}R} =  (\Omega^2 - \Omega_{\rm K}^2) R + g_{\rm m}, \label{eq:rdyn}\\
  &-{1\over 4\pi} {{\rm d}\over {\rm d}R}\!\left[{\mdot\over C_{\rm m}} (\Omega R^2-j_0)\right]  = {{\rm d}\over {\rm d}R}\!\left(\tau_{\phi R}R^2 H\right) + \tau_{\rm m}RH, \label{eq:omegadyn}
\end{eqnarray}
respectively. Here, $j_0$ is the eigenvalue of the system, which represents the specific angular momentum accreted by BH under the no-torque condition (i.e. $\tau_{\phi R}$ = $\tau_{\rm m}$ = 0 just outside of BH horizon, \citealt*{Narayan1998}). In our case, with the boundary conditions given at $R_{\rm out}$, the exact value of $j_0$ is determined numerically, in order to have a smooth transonic solution. In the above equations, the global magnetic field gives a contribution to the pressure, $P_{\rm m}$, decelerates the flow radially, described by $g_{\rm m}$, and exerts a stress/torque to transport the angular momentum, $\tau_{\rm m}$. As detailed in {\it Appendix} \ref{app:madeqs}, the last two terms can be approximately expressed as (e.g., \citealt{Oda2007, LC2009, Cao2011}),
\begin{eqnarray}
g_{\rm m} & \approx & {1\over 4\pi \rho }\left({\br \bz\over H} - {\br^2+\bphi^2\over 4R}\right), \label{eq:gm}\\
\tau_{\rm m} & \approx & {R \over H}\,{\bz\bphi\over 4\pi}, \label{eq:taum}
\end{eqnarray}
where $\br$ and $\bphi$ are defined at the surface ($z=H$; the superscript `s' stands for disk surface) of MAD, while $\bz$ is the field component at the midplane (though we assume it to be a function of $R$ only, see Sec.\ \ref{sec:mag}).

Both the MII-driven (or MRI-driven in case of SANE) turbulence and the global fields transport angular momentum outwards ($\tau_{\phi R}$ and $\tau_{\rm m}$ are negative). The ratio between these two stresses can be expressed as (see Sec.\ \ref{sec:mag} for the definition of the coefficient $\kappa_\phi$)
\begin{eqnarray}
{\tau_{\rm m}\over \tau_{\phi R}} & \approx & {2|\kappa_\phi| \over \alpha (1+\beta_{\rm t}^{-1})}\,{\bz^2/8\pi\over P_{\rm g}} \nonumber\\
& \approx & 0.4\,\left({\alpha\over0.3}\right)^{-1}\left({\beta_z\over5}\right)^{-1} \left({|\kappa_\phi|\over 0.5}\right).
\end{eqnarray}
Here $\beta_z = 8\pi P_{\rm g}/\bz^2$. For typical parameters considered in MAD, MII-driven turbulence and global fields are of comparable importance in the angular momentum transfer. We note that this is different from the recent MHD simulations of \citet{White2019}, where they found that the stress of global field dominates the transfer of angular momentum. This difference is because the effective viscosity parameter of the turbulent stress is $\alpha\sim 0.01$--0.1 in their MHD simulations, while we adopt a larger value of $\alpha=0.3$, as suggested by observations of BHBs in their hard state.

In our model the radial distribution of global field is set in advance (see Section \ref{sec:mag}). Consequently an extremely high magnetic flux (or $\bz$) may lead to ${\rm d}\Omega/{\rm d}R > 0$ and even $\Omega < 0$ in the inner regions. However, in reality, magnetic flux will diffuse outward due to the MII, which, in turn, prevents such high flux accumulation inside $\rmag$. The distribution of the global field inside $\rmag$ would then be adjusted, and the torque $\tau_{\rm m}$ reduced. Considering this effect, we impose ${\rm d}\ln\Omega/{\rm d}\ln R = -1$ for those cases. We note that such correction is only necessary in the innermost regions of high-$\Phi_{\rm n}$ MAD systems.

The density in MAD, even that of the gas spirals, is fairly low. The timescale for electrons and ions to reach thermalization balance is usually longer than the accretion timescale, and the accretion flow will be two-temperature (for SANEs, see \citealt*{Narayan1998, Yuan2014} for related discussions). The gas pressure thus has two components, i.e. we have $P_{\rm g} = P_{\rm i} + P_{\rm e} \equiv \rho k T_{\rm i}/\mu_{\rm i} m_p + \rho k T_{\rm e}/\mu_{\rm e} m_p$, where the effective molecular weight of ions and electrons are respectively, $\mu_{\rm i} = 1.23 $ and $\mu_{\rm e} = 1.14$ (taking the H mass fraction of $0.75$), $k$ is Boltzmann's constant and $m_p$ is the proton mass. We then write the energy equations separately for ions and electrons,
\begin{eqnarray}
\rho V_R \left({d\epsilon_{\rm i}\over dR} - {P_{\rm i}\over \rho^2}{d\rho\over dR}\right) & = & (1-\delta) q_{\rm heat} - q_{\rm ie},\label{eq:ionapp}\\
\rho V_R \left({d\epsilon_{\rm e}\over dR} - {P_{\rm e}\over \rho^2}{d\rho\over dR}\right) & = & \delta q_{\rm heat} + q_{\rm ie} - q_{\rm rad},\label{eq:eapp}
\end{eqnarray}
respectively. Here $\epsilon_{\rm i} = a(\theta_{\rm i})k T_{\rm i}/\mu_i m_p$ and $\epsilon_{\rm e} = a(\theta_{\rm e}) k T_{\rm e}/\mu_e m_p$ are the internal energy (per unit mass) of ions and electrons, respectively, $\theta_{\rm i} = k T_{\rm i}/m_p c^2$ and $\theta_{\rm e} = k T_{\rm e}/m_e c^2$ are the respective dimensionless temperatures, $a(\theta)$ is a function of adiabatic index, with $a(\theta) = 3/2$ for $\theta\ll 1$, $q_{\rm ie}$ is the ion-electron energy transfer rate due to Coulomb interactions \citep{Stepney1983}, $q_{\rm rad}$ is the radiative cooling rate (see Sec.\ \ref{sec:rad} below), $q_{\rm heat}$ is the turbulent heating rate, and $\delta$ defines the fraction of the dissipated energy received by the electrons.

There are several possible dissipation mechanisms to heat electrons and ions in accretion disks (cf.\ \citealt*{Xie2012, Xie2016} for brief summaries). Among them, the two leading models proposed in literature are magnetic reconnection (e.g., \citealt*{Sironi2014, Sironi2015, Numata2015, Rowan2017, Ball2018}) and the Landau-damped MHD turbulent cascades (e.g., \citealt*{QG1999, Howes2010}. In accretion systems, both could happen, irrelevant to what drives the turbulence. In MAD systems outside $\rmag$, the energy dissipation is driven by the conventional MRI, while inside $\rmag$ the turbulence is driven by the MII. Despite this difference, we simply take the turbulent heating rate to be related only to the turbulent stress part \citep{SS1973}, i.e.
\begin{equation}
q_{\rm heat}=\tau_{\phi R} {{\rm d}\Omega\over {\rm d}R}. \label{eq:qheat}
\end{equation}
Then, $\delta$ plays an important role in determining the radiated power \citep{Xie2012}. Observations of BH accretion systems typically require $\delta \sim 0.1$--0.5 (cf. \citealt*{Xie2012} and references therein), while theoretical investigations of both the magnetic reconnection and the damped turbulent cascades agree that $\delta$ can be fairly large, and its value increases with decreasing $\beta$ (e.g., \citealt*{Howes2010, Sironi2014, Sironi2015, Ball2018}). For simplicity, we assume here $\delta$ to be constant.

Finally we note that the equations above are quite general. When the magnetic flux, $\Phi_{\rm n}$, is low and the MII is not triggered, we have $C_{\rm m} = 1$ at all radii and these equations above will naturally describe the SANE system with global fields (e.g., \citealt*{Oda2007, Oda2012, LC2009, Cao2011}). If the global field is absent (i.e., $g_{\rm m} = 0$ and $\tau_{\rm m} = 0$), the equations become those for the standard SANE model (\citealt*{Yuan2014}, see Table \ref{tab:name}).

\subsubsection{Global Magnetic Field Configuration} \label{sec:mag}

Global fields are of crucial importance for the dynamics of MAD. However, their structure cannot be determined from first principles. We thus set the field structure based on the 3D MHD simulations of \citetalias{McKinney2012}. We first assume all the field components have no azimuthal dependence within either the spirals or voids; however, the field strengths are different between them. We now set their distribution in the vertical direction, focusing on properties in the $|z/H| \lesssim 1$ region. We assume $B_z(R, z)$ to be constant along $z$. The radial, $B_R(R, z)$ and azimuthal, $B_\phi(R, z)$, components are assumed to be odd and linear functions of $z$, i.e., $B\propto z$. Thus, the field components are fully determined by $B_z(R)$ at the midplane, and by $\br(R)$ and $\bphi(R)$ at the inflow surface.
% defined by the respective values at the inflow surface.
% B_z, \br and \bphi have already been defined below Equation 6.

We then design the radial distribution of global fields. As long as we initially have a relatively strong poloidal magnetic field, the initial radial distributions of $\br$, $\bphi$ and $\bz$ in the simulations of \citetalias{McKinney2012} are found to be almost irrelevant to the final field configuration, implying a universal field structure in MAD systems. Furthermore, the dominant field component is $\bz(R)$, whose radial distribution follows a power-law form at zeroth order approximation. We thus express $\bar{B}_z(R)$, defined as the strength of $\bz(R)$ azimuthally-averaged over the spirals and voids (see Equation \ref{eq:bav} below), as
\begin{equation}
\bar{B}_z(R) = \bar{B}_{z0} \left({R\over R_0}\right)^{-s_{bz}}, \label{eq:bz}
\end{equation}
where $s_{bz}$ represents the distribution slope, which we assume $= 1.1$ based on the results of \citetalias{McKinney2012}.

We now consider the magnetic fields in the gas spiral regions. We note first that \citet*{Lubow1994} investigated field dragging in geometrically-thin accretion disks, where they found that competition between advection (through viscosity) and diffusion (through magnetic resistivity) is determined by the magnetic Prandtl number, $\pr$. With the assumption that $B_R(R,z)$ is an odd function of $z$ (adopted in this work and also in \citealt*{Lubow1994}), they further found $\br$ to be
\begin{equation}
  \br = {1\over\pr} {H\over R} \bz. \label{eq:br}
\end{equation}
\citet*{Guan2009} found that only when $\pr \ga R/H$ the vertical field can be advected inward by accretion, rather than be diffused outward. Obviously efficient field accumulation/advection is required in order to enter into the MAD phase, this condition should be met in MAD systems. Considering typical values of $H/R\sim 0.3$--0.5 of MAD, we set $\pr = 2$ as the fiducial value. With given $\bz$, we have a relatively weaker $\br$, i.e. $\br \approx 0.2 \bz$. This is consistent with MHD simulations of MAD, cf. \citetalias{McKinney2012}.

The $\phi$-component $\bphi$ directly connects to the rotation of the accreting gas, i.e. the field ratio $\bphi/\br$ correlates positively with $\Omega$ but negatively with $|V_R|$. With Equation\ (\ref{eq:br}) we then write $\bphi$ as,
\begin{equation}
\bphi = \kappa_\phi {\Omega R\over |V_R|+\Omega_{\rm K} R} {H\over R} \bz,\label{eq:bphi}
\end{equation}
where $\kappa_\phi$ is a free parameter. In this expression, the Keplerian velocity $\Omega_{\rm K}R$ is introduced to reduce the $\bphi$ value in case of slower rotation. Also, this form ensures $\bphi$ to be always smaller than the dominant component $\bz$. Since $\br\bphi < 0$, we take a negative value of $\kappa_\phi = -0.5$ as a fiducial value. For a typical value of $\Omega \approx 0.5\,\Omega_{\rm K}$, we will have $\bphi\sim -(0.1$--$0.2)\bz$.

Equations \ref{eq:br} and \ref{eq:bphi} determine the magnetic fields in gas spirals and outside $\rmag$. We now consider the highly-magnetized voids inside $\rmag$. We first assume the radial field component in the voids is negligible. We then estimate the azimuthal component $B_{\phi{\rm,v}}$ through the dynamical coupling between the magnetized voids and the gas spirals. Due to strong magnetic stress in the magnetized voids, the rotational velocity of the voids is lower by a factor of $\sim$0.5--0.9 than that of their neighboring (in azimuthal direction) gas spirals (\citetalias{McKinney2012}). We thus assume the azimuthal-to-vertical field ratio following the relationship, 
\begin{equation}
{B_{\phi{\rm,v}}^{\rm s}\over B_{z{\rm,v}}} \approx 0.8 {B_{\phi{\rm,s}}^{\rm s}\over B_{z{\rm,s}}},
\end{equation}
where the factor of 0.8 is set arbitrarily.

The main quantity to describe MAD systems is the magnetic flux. The half-surface magnetic flux, $\Phi (R)$, within a radius $R$ is
\begin{equation}
\Phi (R) = 2\pi\int_{2 \rg}^R R\, \bar{B}_z(R)\,dR. 
\end{equation}
The dimensionless magnetic flux normalized by the mass accretion rate has also been often used (\citealt*{Tch2011}; for alternative definitions that differ by a constant factor, see \citealt*{Penna2010}; \citetalias{McKinney2012}),
\begin{equation}
\Phi_{\rm n}(R) = {\Phi(R)\over \left[\dot{M}(R)R_{\rm g}^2 c\right]^{1/2}},
\end{equation}
where the subscript `n' denotes a normalized value. We note that here we only consider the magnetic flux threading onto the accretion disk, but not that threading onto the BH horizon, $\Phi^{\rm BH}_{\rm n}$, which is integrated over one BH hemisphere. That quantity determines the power of the jet lauched via the mechamism of \citet{BZ77}, see \citet{Tch2015}, but likely it has only a weak impact on the accretion flow. MHD simulations show that the maximal/saturated value of $\Phi^{\rm BH}_{\rm n}$ depends on the disc thickness, namely $\Phi^{\rm BH}_{\rm n, max} \approx 50\,[(H/R)/ 0.3]$, see e.g., \citet*{Tch2011}; \citetalias{McKinney2012}.

\subsubsection{Criteria for the Magnetic Interchange Instability and the Evaluation of $\rmag$}\label{sec:rmag}

We here determine the location $\rmag$. We have two conditions. One is that the total magnetic stress force $g_{\rm m,all}$ (including the magnetic pressure) is strong enough to complete against the effective gravity \citepalias{Narayan2003}, namely
\begin{equation}
g_{\rm m,all}\approx {\bar{B}_z^2\over 4\pi \Sigma} > f_g\, g_{\rm eff},\label{eq:grav.vs.b}
\end{equation}
where $g_{\rm eff} = (\Omega_{\rm K}^2-\Omega^2)R$ is the effective gravity (per unit mass). Factor $f_g\approx 0.5$ is introduced (admittedly arbitrarily) to compensate for the strong outward force due to (gas+turbulent field) pressure. This provides a more detailed expression compared to Equation (\ref{R_m}), where neither the pressure gradient nor the centrifugal force due to rotation are considered. With $\Omega/\Omega_{\rm K}\approx 0.3$--0.5, we have $f_\Omega=1-\Omega^2/\Omega_{\rm K}^2\approx 0.75$--0.9. After some algebraic manipulations, we have (in case of outflow,   $\dot{M}$ represents the accretion rate at $\rmag$) 
\begin{equation}
 {\rmag\over \rg} \approx \pi^{-\frac{4}{3}}\! \left({\alpha\over f_g f_\Omega}\right)^{\frac{2}{3}}\! \left(R_{\rm g}^2 c\right)^{-\frac{2}{3}} \dot{M}^{-\frac{2}{3}}\Phi^{\frac{4}{3}}\left({H\over R}\right)^{\frac{4}{3}}\!.
\label{eq:rmag_detail}
\end{equation}
The difference in the dependence on $H/R$ appears because the pressure of global field is now taken into account, i.e. in Equation (\ref{eq:grav.vs.b}) we adopted $\bz^2$ instead of $\br\bz$ (see Equations \ref{eq:rdyn} and \ref{eq:gm}).

The other condition that we adopt is that for appearance of the MII. Analyses of the disk instability in a thin disk threaded by a vertical field $\bz$ show that this requires that $\Sigma/\bz$ increases with decreasing $R$ fast enough to overcome the stabilizing effect of the velocity shear \citep{Spruit1995, Kulkarni2008}, namely 
\begin{equation}
\gamma_{\rm B\Sigma}^2 \equiv g_{\rm eff} {d\over dR}\left(\ln \left|{\Sigma\over \bz}\right|\right) \geq 2 \left(R{d|\Omega|\over dR}\right)^2 \equiv \gamma_\Omega^2.
\end{equation}
The location of $\rmag$ is determined by applying both criteria.

\subsubsection{Connections between Gas Spirals and Magnetized Voids}\label{sec:coupling}

In our model, all the mass are in the gas spirals and the magnetized voids are assumed to be gas-free. Meanwhile, although spatially separated, the spirals and the voids are coupled together, e.g., the magnetized voids roughly corotate with the gas spirals \citepalias{McKinney2012}, although the rotational velocity of the magnetized voids, whose magnetic field tension is much stronger, is slower than that of gas spirals. We assume that the spirals and the voids reach a pressure equilibrium state. Since the radial field component is negligible in the magnetized voids, we can express the pressure balance condition ($P_{\rm tot,v} = P_{\rm tot,s}$) as 
\begin{eqnarray}
{{B_{z{\rm,v}}}^2 + {B_{\phi{\rm,v}}^s}^2/4\over 8\pi} & = & P_{\rm g,s} + P_{\rm tm,s} + \nonumber\\
  & & {{B_{z{\rm,s}}}^2+({B_{R{\rm,s}}^{\rm s}}^2+{B_{\phi{\rm,s}}^{\rm s}}^2)/4\over 8\pi}.\label{eq:equalp}
\end{eqnarray}
Here factor $1/4$ is introduced to account for the vertical averaging of both $\br(R,z)$ and $\bphi(R,z)$ (see Sec.\ \ref{sec:mag}).

Note that using the definition of the total plasma parameter of the spirals as $\beta_{\rm s} = P_{\rm g,s}/(P_{\rm tm,s}+P_{\rm m,s})$, the pressure balance also provides an estimate of the field strength ratio between spirals and voids, 
\begin{equation}
{B_{z{\rm,v}}\over B_{z{\rm,s}}}\approx \left({P_{\rm tot,s}\over P_{\rm m,s}}\right)^{1/2} = \left({1+\beta_{\rm s}\over 1-\beta_{\rm s}/\beta_{\rm t}}\right)^{1/2}. \label{eq:bz_void_to_spiral}
\end{equation}
For the assumed $\beta_{\rm t}=10$, we have $B_{z{\rm,v}}/B_{z{\rm,s}}\approx 3.5$, 5.2 for $\beta_{\rm s} = 5$, 7, respectively. Physically, it corresponds to the expelling of a significant fraction of the magnetic fields out from the gas spirals into the magnetized voids.

\subsubsection{Transition Conditions at $\rmag$ and Covering Factor $C_{\rm m}$}\label{sec:spiralmodel}

The following transition conditions at $\rmag$ are adopted. Basically, we re-arrange the magnetic fields and the gas density, but keep the velocities ($V_R$ and $\Omega$) and temperatures ($T_{\rm e}$ and $T_{\rm i}$) unaffected, 
\begin{eqnarray} 
  V_{R,{\rm s}} = V_R, \hspace{0.3cm} \Omega_{\rm s} = \Omega, \nonumber\\
  T_{\rm e,s} = T_{\rm e}, \hspace{0.3cm} T_{\rm i,s} = T_{\rm i}.
\end{eqnarray}
This means that the heating (increasing $T_{\rm i}$ and $T_{\rm e}$ in the spirals) during the compression in the $\phi$ direction is neglected. For the gas density, we iteratively adjust the guessed density in the spirals at $\rmag$ until the accretion rate at $\rmag-$ converges to that at $\rmag+$. Thus, we assume quasi-steady state without any mass accumulation/pile-up outside of $\rmag$, which corresponds to the results of 3D simulation where MII is observed to be triggered automatically (\citetalias{McKinney2012}; \citealt{White2019}).

We take the following two conditions for the transition of the global magnetic fields. First, a pressure balance between the dense spirals and the magnetized voids is roughly achieved, see Sec.\ \ref{sec:coupling}. Second, when averaged in the azimuthal direction, the magnetic field in this nonaxisymmetric region still follows the $R^{-s_{bz}}$ profile, see Equation\ (\ref{eq:bz}). Specifically, we have,
\begin{eqnarray}
  \bar{B}_z& = C_{\rm m}\, B_{z{\rm,s}} + (1 - C_{\rm m})\, B_{z{\rm,v}}.\label{eq:bav}
\end{eqnarray}
From Equations\ (\ref{eq:equalp}) and (\ref{eq:bav}), one immediate result is that $B_{z{\rm,s}} < \bar{B}_z$ while $B_{z{\rm,v}} > \bar{B}_z$ (cf. {\it Appendix} \ref{app:cmin} for details), i.e.\ magnetic fields are expelled out into the highly magnetized voids. In this case, the spirals can still remain weakly-magnetized (with $\beta_{\rm s} > 1$) even when the whole system has a large magnetic flux, which is indeed the case as observed in numerical simulations, see \citetalias{McKinney2012}.

Although the exact value of $C_{\rm m}$ cannot be determined from basic physics, there exist two lower limit constraints. One is from our treatment of the radiative processes (see Sec.\ \ref{sec:rad}). We assume the emission from the spirals is quasi-isotropic, i.e., side emission equals to surface emission. This is valid only when the azimuthal length of each spiral is comparable to its thickness ,$H$. There is then the corresponding constraint on $C_{\rm m}$ (cf.\ Equation\ \ref{eq:ns}), with the number of spirals greater than unity, i.e.\ $n_{\rm s} > 1$,
\begin{equation}
C_{\rm m} > {H/R\over\pi},\label{eq:cmin2}
\end{equation}
which is approximately in the range of 1/10--1/4. In our calculations, we assume 
$C_{\rm m} > 1/5$.

The other constraint comes from the spiral-void coupling. Through an analysis of Equations\ (\ref{eq:equalp}) and (\ref{eq:bav}), we find a lower limit on $C_{\rm m}$, below which there is no pressure-balance solution (see {\it Appendix} \ref{app:cmin} for the derivation), namely
\begin{equation}
C_{\rm m} > C_{\rm m, min}\equiv {\rm max}\left[0,\hspace{0.2cm} 1- \left({1+\beta_{\rm t}\over f_3\,\beta_{\rm t}} {P_{\rm g,s}\over \bar{B}_z^2/8\pi}\right)^{-\frac{1}{2}}\right],\label{eq:cmin}
\end{equation}
where $f_3 = 1 + {B_{\phi{\rm,v}}^{\rm s}}^2/(4B_{z{\rm,v}}^2)$ (cf.\ {\it Appendix} \ref{app:cmin}). Thus, stronger magnetic fields allow for narrower occupation in the azimuthal direction (smaller $C_{\rm m, min}$) for the gas spirals. 

We have found no method to determine the actual value of $C_{\rm m}$. Instead, we simply use the lower limits, which correspond to the magnetic pressure from the voids exerted on the spirals being strong. We combine the above two constraints and arbitrarily set $C_{\rm m}$ as 
\begin{equation}
C_{\rm m} = (4 C_{\rm m,min} + 1)/5. \label{eq:c_m}
\end{equation}
We note that smaller $C_{\rm m}$ leads to weaker magnetic field (higher $\beta_{\rm s}$) in the spirals.

\subsection{Radiation in MAD Systems} \label{sec:rad}

For the hot version of MAD, we consider the following radiative processes, i.e. synchrotron, bremsstrahlung and the inverse Compton scattering. Note that these processes are those also included in SANE systems \citep{Narayan1995}.

In this work, we focused on hot accretion flow itself. In this case, we do not take into account the emission related to the cold SSD outside of $R_{\rm out}$, i.e. the Comptonization of thermal blackbody emission from SSD (which appears to be a major parameter in accreting BHs, e.g. from the reflection-index correlation, \citealt{zdziarski99}), and the reflection and reprocessing by the cold SSD. Additionally, for electrons in hot accretion flow, we also ignore the possibility that the thermal relativistic-Maxwellian distribution has a weak high-energy tail, which can enhance the synchrotron emission by orders of magnitude, e.g., \citet{veledina11}.

%We consider both the gravitional redshift and relativistic Doppler shifts due to gas motions. We assume the accretion system is face-on and following \citet{Manmoto1997} to define the combined redshift $z_{\rm r}$ as,
%\begin{equation}
%(1+z_{\rm r})^{-1} \equiv {\nu_o\over\nu_e} =\left(1-{2\rg\over R}\right)^{1/2}\,\left(1-{V^2\over c^2}\right)^{1/2},
%\end{equation}

%thus the gas velocity is perpendicular to the line of sight, i.e. $\theta_{\rm v.LOS} = 0$. With the lorentz factor of accreting gas $\Gamma = 1/(1-V^2/c^2)^{1/2}$, we define the combined redshift $z_{\rm r}$ as,
%\begin{equation}
%(1+z_{\rm r})^{-1} \equiv \nu_o/\nu_e = {(1-2\rg/R)^{1/2}\over \Gamma(1-V/c*\cos\theta_{\rm v.LOS})}, 
%\end{equation}
%where $\nu_e$ and $\nu_o$ are respectively the emitted (in local rest frame) and the observed (by distant observer) photon frequencies.

\subsubsection{Radiation outside of $R_{\rm m}$}

We first investigate the region outside of $R>\rmag$. For this region, there is no significant difference between MAD and SANE, as discussed in Sec. \ref{sec:madsim}. The only difference is to take the global magnetic fields into account for the synchrotron emission. We thus directly follow \citet*{Manmoto1997} to calculate them, with emissivities taken from \citet*{Narayan1995}. For the radiative transfer along the vertical direction, we take the Eddington approximation and adopt the radiative diffusion description. As detailed in \citet{Manmoto1997}, the radiative transfer along $z$ direction is then solved under the two-stream approximation \citep{Rybicki1979}. The radiative flux of synchrotron and bremsstrahlung emission at given frequency $\nu$ (in local rest frame), defined as $F_\nu^{\rm seed}$ since it will also serve as seed photons for the Compton scattering process, can then be expressed as \citep{Manmoto1997},
\begin{equation}
F_\nu^{\rm seed} = {2\pi\over \sqrt{3}} B_\nu [1-\exp(-2\sqrt{3}\tau_\nu)].
\end{equation} 
Here $B_\nu$ is the blackbody emissivity per unit solid angle, and $\tau_\nu = \sqrt{\pi/2}\,\kappa_\nu H$ is the half optical depth in vertical direction, from the equatorial plane to the surface. The absorption coefficient $\kappa_\nu$ is defined as $\kappa_\nu = j_\nu/B_\nu$, where $j_\nu$ is the emissivity coefficient that includes both synchrotron and bremsstrahlung. 

We now consider the Compton scattering. In principle, the Compton scattering in hot optically-thin flows happens globally, i.e. seed photons from one location can propagate to a distant location and scatter with energetic electrons there (e.g., \citealt*{Yuan2009, Xie2010, Niedzwiecki2012}, and references therein). For technical reasons we limit ourselves to the local one-zone (in vertical direction) treatment, which is also widely adopted in literature.

The Compton scattering is calculated in two steps, with different objectives. The first is during the procedure of solving the dynamical equations in Sec. \ref{sec:maddyn} and the aim is to determine the dynamical structure of MAD. Here only the radiative cooling rate $q_{\rm rad}$ is required, while the spectral information is not necessary. We adopt the enhancement factor method (\citealt*{Dermer1991}, coefficients for the disk model with seed photon energy fixed to 1 eV.), which provides the ``averaged'' energy boost of seed photons. Once the dynamics is determined, we move to the second step, i.e.\ to derive the spectrum. For this step we take a more accurate treatment on the Compton scattering process, following \citet*{Coppi1990}. 

Once the Compton scattering is done, we can easily sum up the emission from all radiative mechanisms to derive the total radiative flux $F_\nu$ at each radius $R$. With the definition of the frequency-integrated radiative flux per unit area $F_{\rm rad}$ ($\equiv \int F_\nu d\nu$), the local radiative cooling rate (per unit surface area $Q_{\rm rad}$ and per unit volume $q_{\rm rad}$) can then be evaluated as
\begin{equation}
Q_{\rm rad} = 2F_{\rm rad}, \hspace{0.7cm} q_{\rm rad} = {F_{\rm rad}\over H}.
\end{equation}
Both are identical to those of SANEs (e.g. \citealt*{Narayan1995, Manmoto1997}). The radiative cooling $q_{\rm rad}$ determines the energy balance for electrons (cf. Equation \ref{eq:eapp}). Moreover, it can also indirectly impact on the dynamical structure of the flow (cf. the energy equation for ions Equation \ref{eq:ionapp}, through electron-ion coupling $q_{\rm ie}$) when the accretion rate is high and $q_{\rm rad}$ is comparable to the turbulent heating $q_{\rm heat}$.

Finally, through radial integration, the total luminosity (measured by distant observer) $L_{\nu_o}$ at observed frequency $\nu_o$ is given by \citep{Ghisellini2013}
\begin{equation}
L_{\nu_o} = 2\int {2\pi R\,F_{\nu}\over (1+z_{\rm r})^3}\,dR = 4\pi\int {R\,F_{\nu}\over (1+z_{\rm r})^3}\,dR. \label{eq:lum}
\end{equation} 
Here, the combined redshift $z_{\rm r}$, defined as $(1+z_{\rm r})^{-1} = \nu_o/\nu$, includes both the gravitational redshift and relativistic Doppler shift. The additional factor $2$ accounts for the two surfaces of the accretion flow. 

\subsubsection{Radiation inside of $R_{\rm m}$} \label{sec:rad.inside.rm}

The nonaxisymmetric region (inside $\rmag$) deserves additional consideration. We start from examining the relative importance of the emission from the highly-magnetized voids. For this purpose, we temporarily abandon the gas-free approximation but consider the residual dilute gases in the voids. We take from numerical simulations (e.g. \citetalias{McKinney2012}) a plasma parameter value $\beta_{\rm v}\sim 0.01$ for the magnetized voids. For the synchrotron emission, we assume the electrons are relativistic and estimate the void-to-spiral luminosity radio at each ring of radius $R$ as (see Equation\ 6.7 in \citealt*{Rybicki1979}),
\begin{eqnarray}
  {L_{\rm syn,v}(R)\over L_{\rm syn,s}(R)} & \approx & {\rho_{\rm v}\over\rho_{\rm s}}\, {B_{\rm v}^2\over B_{\rm s}^2}\, {T_{\rm e,v}^2\over T_{\rm e,s}^2}\, {1-C_{\rm m}\over C_{\rm m}} \nonumber\\
  & \approx & \left({T_{\rm i,v}\over T_{\rm i,s}}\right)^{-1}\,\left({T_{\rm e,v}\over T_{\rm e,s}}\right)^{2}\,{\beta_{\rm v}/(1+\beta_{\rm v})^2\over \beta_{\rm s}/(1+\beta_{\rm s})^2}\, {1-C_{\rm m}\over C_{\rm m}}\nonumber\\
  &\approx& {\beta_{\rm v}\over \beta_{\rm s}/(1+\beta_{\rm s})^2}\, {1-C_{\rm m}\over C_{\rm m}},\label{eq:syn_void_spiral}
\end{eqnarray}
where we consider the fact that the spirals and voids are in pressure balance and that $\rho T_{\rm i} B^2 \propto P_{\rm g}(P_{\rm tm}+P_{\rm m})\propto P_{\rm tot}^2\times \beta/(1+\beta)^2\propto\beta/(1+\beta)^2$. In the final expression of Equation (\ref{eq:syn_void_spiral}), we further assume $T_{\rm e,v}\approx T_{\rm e,s}$ and $T_{\rm i,v}\approx T_{\rm i,s}$. We thus have $L_{\rm syn,v}(R)/L_{\rm syn,s}(R)\approx 0.1$ for $\beta_{\rm s}=5$ and $C_{\rm m}=0.4$, i.e. synchrotron emission from the residual dilute gases in the voids plays a secondary role. In addition, due to the strong dependence on optical depth (proportional to gas density), the Compton scattering in the magnetized voids, either internal or external (e.g., seed photons are from the synchrotron emission of the spirals), is also unimportant. We thus conclude that emission from the voids can safely be neglected, and the gas-free approximation is justified.

We then estimate the number of dense spirals $n_{\rm s}$ at each radius. For simplicity, we assume the width of each spiral is comparable to the total thickness of the disk ($\sim 2H$). The number of spirals can then be estimated as,
\begin{equation}
n_{\rm s} \approx {2 \pi R C_{\rm m}\over 2 H} = \pi C_{\rm m}\left({H\over R}\right)^{-1}. \label{eq:ns}
\end{equation}
For typical values of MAD ($H/R\approx0.3$--0.5 and $C_{\rm m}\approx0.3$--0.5), we have $n_{\rm s}\approx 1$--3. It is in good agreement with simulations where one or two dense spirals dominant (see, e.g., \citealt*{Tch2011}; \citetalias{McKinney2012}). We note that recent simulations with even higher resolutions incline to favor  smaller but more numerous dense blobs \citep{White2019}. We devote further refinement to future work.

Below we derive the emission from gas spirals. Compared to SANEs, the spirals have four surfaces (bottom, top, and additionally two sides neighboring to the magnetized voids). Since the width of each spiral is assumed to be comparable to $2H$, the optical depth in azimuthal direction is roughly the same to that in vertical direction. Under this condition, the emission of the gas spirals can be consistently approximated as nearly isotropic, i.e. radiation (per unit area) from each side equals to that from each surface. With Equation (\ref{eq:ns}) we thus have
\begin{eqnarray}
Q_{\rm rad} & \approx & {(F_{\rm rad}\,4\pi C_{\rm m}RdR + F_{\rm rad}\,2n_{\rm s}\,2HdR)\over 2\pi C_{\rm m} RdR}\nonumber\\
& = & 4 F_{\rm rad}.
\end{eqnarray}

\begin{figure*}
\centering
\includegraphics[width=15.cm]{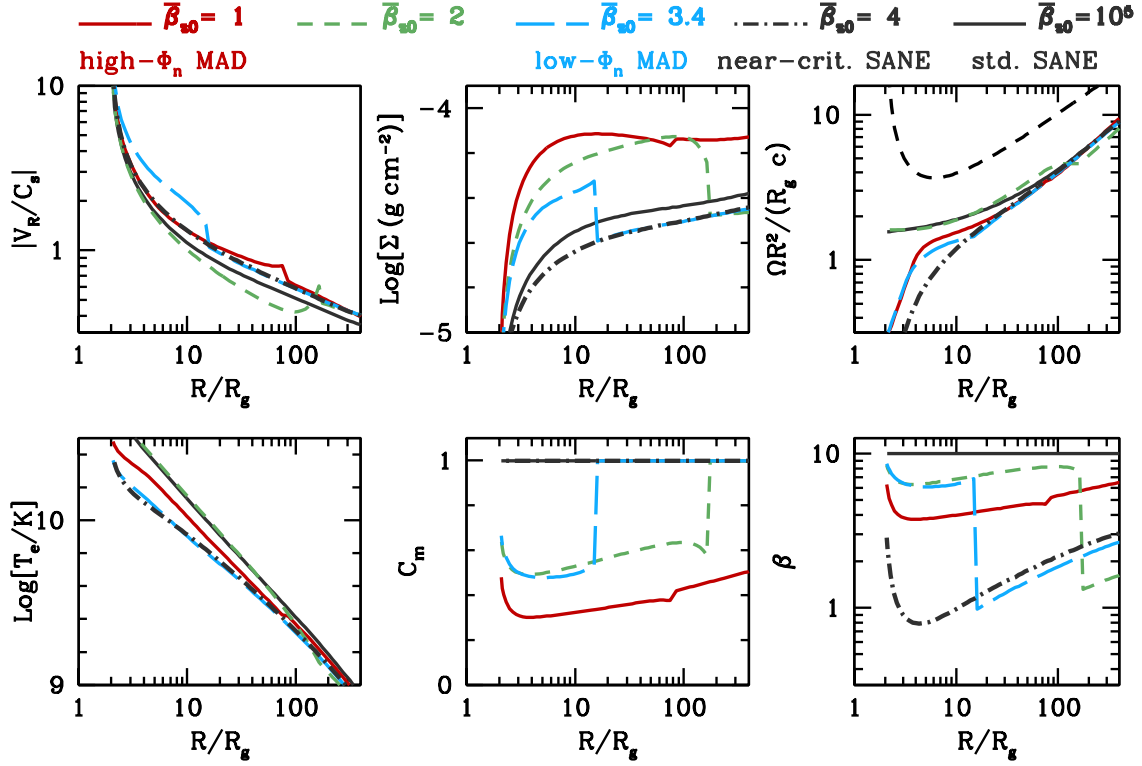}
\caption{Dynamical structure of MAD at low accretion rate ($\dot{M}_0 = 10^{-5}\ \dot{M}_{\rm Edd}$ at $R_{\rm out}=10^3 R_g$, the outflow parameter $s=0.3$). From left to right, the top panels show the Mach number $V_R/c_{\rm s}$, the half surface density $\Sigma$, and the specific angular momentum $\Omega R^2$; the bottom panels show the electron temperature $T_{\rm e}$, surface covering factor $C_{\rm m}$ and the plasma $\beta$-parameter in gas regions. As listed in Table \ref{tab:par}, basic parameters for global fields are $s_{bz}=1.1$, $\pr = 2$, $\kappa_\phi = -0.5$, and others related to microphysics are $\alpha=0.3, \beta_{\rm t}=10, \delta=0.1$. The strength of $\bz$ is set through $\betazrout$ at $R_{\rm out}$, where $\betazrout = 1$ (red solid; high-$\Phi_{\rm n}$ MAD with $\rmag > 10^3 \rg$), $=2$ (green dashed, MAD with $\rmag \approx 160 \rg$) and $=3.4$ (blue long-dashed; low-$\Phi_{\rm n}$ MAD with $\rmag \approx 15 \rg$). The transition to the spiral/void region is evidently shown in panels of $\Sigma$, $\beta_{\rm s}$ and $C_{\rm m}$. For comparison, we also show the standard ($\betazrout=10^5$) and near-critical ($\betazrout=4$) SANEs as respectively, the black solid and black dot-dashed curves. The upper black dashed curve in $\Omega R^2$ plot represents the Keplerian rotation.}
\label{fig:dynlowmdot}
\end{figure*}

The radiative cooling rate $q_{\rm rad}$ can then be expressed as
\begin{equation}
q_{\rm rad} \approx 2 {F_{\rm rad}\over H}.
\end{equation}
The total luminosity (measured by distant observer) in this region can now be derived as, 
\begin{eqnarray}
L_{\nu_o} & \approx & \int {(F_{\nu}\,4\pi C_{\rm m}RdR + F_{\nu} \,2n_{\rm s}\,2HdR)\over (1+z_{\rm r})^3} \nonumber\\
&\approx & 8\pi \int {C_{\rm m}\,R\,F_{\nu}\over (1+z_{\rm r})^3}\, dR. \label{eq:lumspiral}\
\end{eqnarray}
We caution that the apparent factor of 2 enhancement in Equation\ (\ref{eq:lumspiral}) compared to Equation (\ref{eq:lum}) is an artifact that reflects our simplification. In reality dense spirals will irradiate each other. Consequently at least part of the side emission cannot propagate to infinity.

\setlength{\tabcolsep}{3pt}
\begin{table}
\centering
\caption{Model Parameters}
\begin{tabular}{ccl}
\hline
Parameter & Value & Definition/Note\\
\hline
\multicolumn{3}{c}{\it Basic Parameters (fixed in this work)}\\
\hline
$\mbh$ & $10^8$ & Black hole mass (in unit $\msun$)\\
$s$ & $0.3$ & Outflow parameter, $\dot{M}(R)\propto R^s$\\
$\alpha$ & $0.3$ & Viscosity parameter\\
$\beta_{\rm t}$ & $10$ & Gas-to-turbulent-magnetic pressure ratio\\
$\delta$ & $0.1$ & Fraction of electron viscous heating\\
$s_{bz}$ & 1.1 & Radial distribution of $\bar{B}_z$, $\bar{B}_z\propto R^{-s_{bz}}$\\
$\pr$ & 2 & Magnetic Prandtl number\\
$\kappa_\phi$ & -0.5 & Parameter for $\bphi/\bz$. \\
& & $\kappa_\phi<0$, since we have $\br\bphi < 0$\\
\hline
\multicolumn{3}{c}{\it Boundary Conditions (variables in this work)}\\
\hline
$R_{\rm out}/R_{\rm g}$ & -- & Outer boundary of hot accretion flow\\
$\mdot_0/\medd$ & -- & Mass accretion rate at $R_{\rm out}$\\
$\betazrout$ & -- & Averaged plasma $\beta$-parameter at $R_{\rm out}$, \\
& & defined as $\betazrout = 8\pi C_{\rm m} P_{\rm g}/\bar{B}_z^2$\\
\hline
\end{tabular}\label{tab:par}
\end{table}

\subsection{Model Parameters}\label{sec:modelpar}

We summarize in Table \ref{tab:par} all the parameters in our MAD model. We set the BH mass to $\mbh = 1\times10^8\msun$. We additionally fix several parameters of MAD systems, i.e. $\alpha=0.3$, $\beta_{\rm t}=10$, $\delta=0.1$, and $s=0.3$. All these parameters are typical for SANE systems. The outer boundary is set to $R_{\rm out} = 10^3 \rg$. There are two boundary conditions at $R_{\rm out}$. One is the mass accretion rate $\dot{M}_0$ (the net accretion rate onto BH is $\dot{M}_{\rm BH} = 0.155\dot{M}_0$ for the values of $s$ and $R_{\rm out}$ assumed here), and the other is the azimuthally-averaged plasma parameter $\betazrout$, which is defined through $\betazrout = 8\pi\,C_{\rm m}\,P_{\rm g}/\bar{B}_z^2$. Note that if the global field is relatively weak (i.e. $\rmag < R_{\rm out}$), the system will remains nearly axisymmetric at the outer boundary, with $C_{\rm m}=1$ and $\bz=\bar{B}_z$ there. In this case, parameter $\betazrout$ returns to its normal physical meaning.

One additional parameter is the radial distribution of the vertical magnetic field $\bar{B}_z$. We take a power-law profile with a slope $s_{bz} = 1.1$ (see Equation \ref{eq:bz} and \citetalias{McKinney2012}). The radial and azimuthal field components are then set by two additional parameters, i.e. $\pr = 2$ and $\kappa_\phi = -0.5$.

In this work there are only two variables, i.e. $\betazrout$ and the accretion rate $\dot{M}_0$ at the boundary. We use them to examine the impact of global fields and/or accretion rate on the dynamical and radiative properties of hot accretion flows.
%In the following section we will vary only two properties at the boundary, i.e. $\betazrout$ and the accretion rate $\dot{M}_0$, to examine the impact of global fields and/or accretion rate on the dynamical and radiative properties of hot accretion flows.

\section{Results}\label{sec:result}

\begin{figure*}
  \centering
  \includegraphics[height=6.5cm]{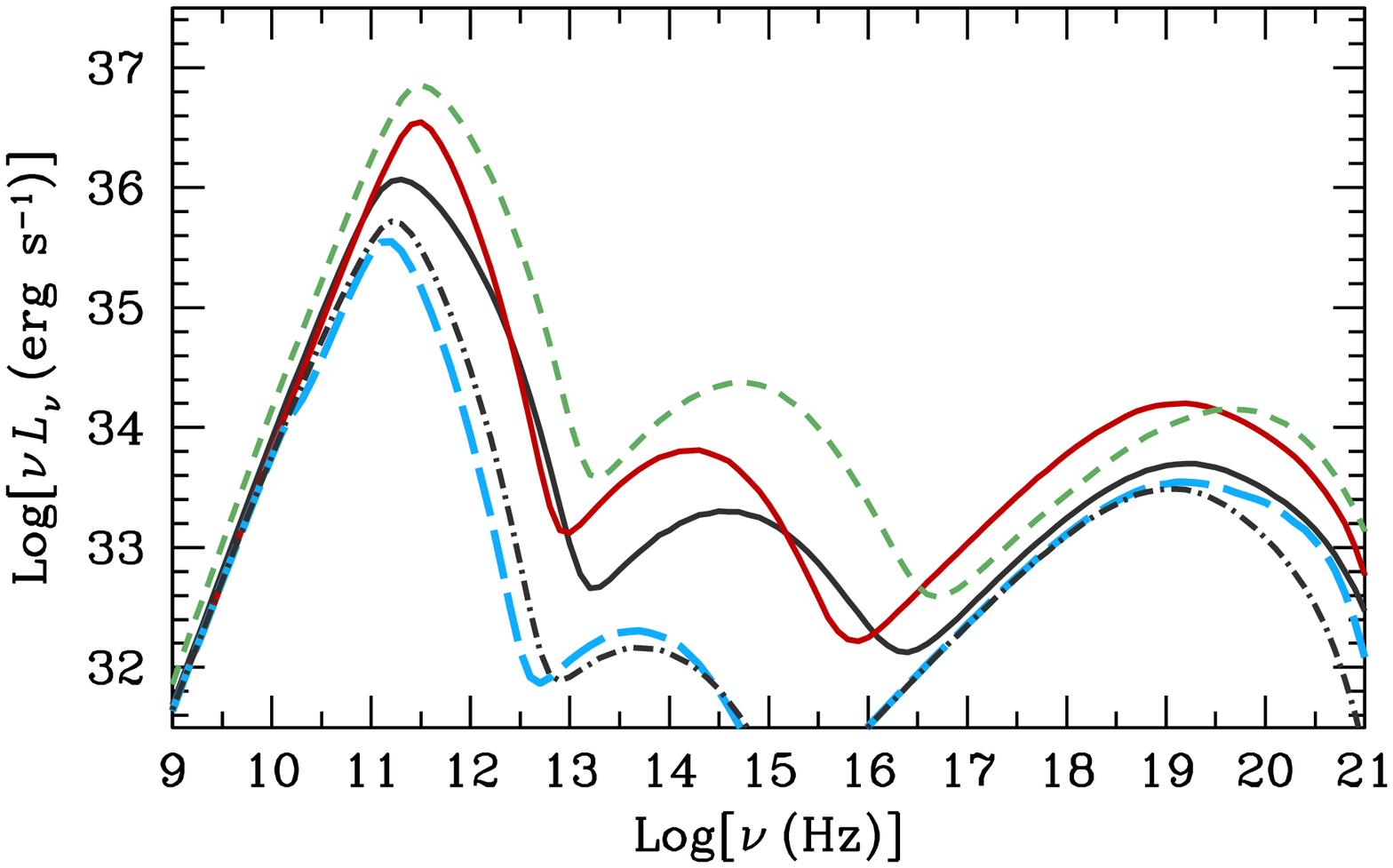} \hspace{1cm}
  \includegraphics[height=6.5cm]{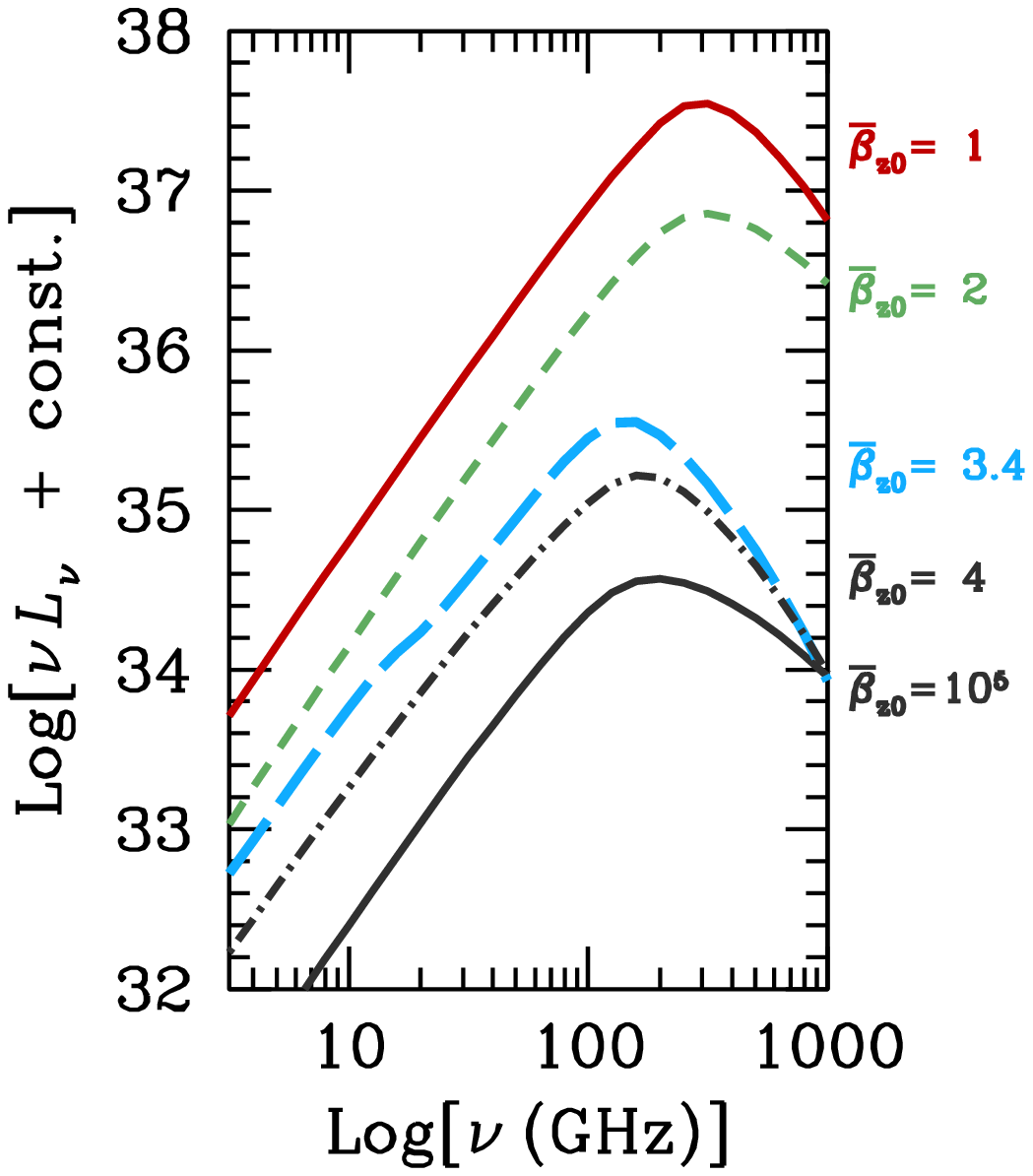}
  \caption{Spectral density distribution (SED) of MAD systems at low accretion rate. As labelled in the {\it Right Panel}, the curves are of the same meaning to those of Figure \ref{fig:dynlowmdot}. The three bumps of each spectrum in the {\it Left panel} are produced by respectively, synchrotron, first-order Compton scattering, and bremsstrahlung. High-order Compton scattering is not important at such low accretion rate. {\it Right panel}: a zoom-in view of the synchrotron bump (flux arbitrarily shifted) in all models.}\label{fig:sedlowmdot}
\end{figure*}

\subsection{Dynamical and Radiative Properties at Low-$\dot{M}/\medd$}\label{sec:lowmdot}

We first examine the typical hot accretion cases, where the accretion rate is sufficiently low and radiative cooling plays a negligible role on the flow dynamics. We here take $\dot{M}_0 = 10^{-5}\medd$ as a representative. Under this accretion rate, the radiative cooling is found to be less than $1\%$ the turbulent heating. For our setup, we find that the system enters into MAD when $\betazrout \lesssim 3.7$--3.8. Obviously, this value is sensitive to the radial distribution of global fields (i.e. $s_{bz}$).

Figure\ \ref{fig:dynlowmdot} shows the dynamical structure for those low-$\dot{M}/\medd$ systems. From left to right, the top panels show the radial distribution of the Mach number ($|V_R/c_{\rm s}|$), the surface density ($\Sigma$) and the angular momentum ($\Omega R^2$), while the bottom panels show that of the electron temperature ($T_{\rm e}$), the covering factor ($C_{\rm m}$) and the plasma $\beta$-parameter of the gas (dense spirals for the spiral/void region). In this plot, the red solid curves are for $\betazrout = 1$, which is the high-$\Phi_{\rm n}$ MAD with $\rmag \ga R_{\rm out} (=10^3 \rg)$ and the dimensionless magnetic flux within $10\rg$ to be $\Phi_{\rm n}(10\rg) \approx 51.6$. The green dashed curves are for $\betazrout = 2$ (MAD with $\rmag \approx 160 \rg$ and $\Phi_{\rm n}(10\rg) \approx 33.6$), and the blue long-dashed curves are for $\betazrout = 3.4$ (low-$\Phi_{\rm n}$ MAD with $\rmag \approx 15 \rg$ and $\Phi_{\rm n}(10\rg) \approx 26.7$). The transition to the spiral/void region is evidently shown by the sharp jumps in the panels of $\Sigma$, $\beta$ and $C_{\rm m}$. For comparison, we also show dynamical structure of the standard ($\betazrout=10^5$) and near-critical ($\betazrout=4$ and $\Phi_{\rm n}(10\rg) \approx 24.8$) SANEs respectively as the black solid and dot-dashed curves.

As clearly demonstrated in Figure \ref{fig:dynlowmdot}, the innermost regions will first transit into MAD phase, if the magnetic flux increases gradually (i.e., with a decrease in $\betazrout$). The transition radius $\rmag$ increases with increasing magnetic flux. For different magnetic fluxes, the dynamical structure of accretion flow outside of $\rmag$ remains similar. This is most evident when comparing the the low-$\Phi_{\rm n}$ MAD (i.e. the blue long-dashed curve) with the near-critical SANE (the black dot-dashed). When MII is triggered and the system enters into the spiral/void regions, we will observe an increase in both the density and the surface density $\Sigma$ of gas spirals (and a decrease of $C_{\rm m}$. Note that $\Sigma \propto -\dot{M}/(V_R C_{\rm m})$.), compared to the SANE cases. For a typical value of $C_{\rm m}\approx0.3$--0.5, we find a factor of 2--3 increase in $\Sigma$. As magnetic flux increases even further, the global fields will force the gas spirals to shrink azimuthally (reducing $C_{\rm m}$). Consequently we will observe a further enhancement in $\Sigma$. Due to additional angular momentum transport by global field, $\Omega$ near BH is considerably small in high-$\Phi_{\rm n}$ MAD systems. We note that the near-critical SANE one, whose magnetic fields within accreting gas are also fairly strong (see the plot of $\beta$), also have rather low angular momentum. 

%The low-$\Phi_{\rm n}$ MAD system, on the other hand, has relatively larger $\Omega$, mostly because of weaker global fields within the spirals, i.e. its $\Omega $ near BH will be more close to that of standard SANE system (not shown here).

The bottom right panel of Figure \ref{fig:dynlowmdot} shows the plasma $\beta$-parameter for the gas regions (i.e. inside of $\rmag$ only that of gas spirals). Outside of $\rmag$, more input of magnetic flux to the whole system results in lower $\beta$ values. Inside of $\rmag$, on the other hand, an opposite effect is observed, i.e.\ more magnetic flux to the whole system will result in larger $\beta$ values (equivalently, relatively weaker magnetic field strength) in gas spirals. This is because in this case more fraction of the global fields are actually expelled out of the dense spirals into the magnetized voids (see Section \ref{sec:coupling} and Equation \ref{eq:bz_void_to_spiral}). The spirals thus have a relatively weaker magnetic field, with $\beta_{\rm s}\approx3$--8. We note that this value is lower than that of the near-critical SANE, whose $\beta$ reaches $\la$1 at innermost regions. One consequence of the expel of magnetic fields out of the gas spirals is that the spirals in MAD systems have a slightly lower scale height (or equivalently aspect ratio $H/R$, not shown here) compared to that of the near-critical SANE. We also observe that MAD with $\betazrout = 2$ has a different dynamics compared to MAD with $\betazrout = 1$, i.e. hotter temperature, lower radial velocity. Correspondingly, it has similar surface density even through its surface covering factor is larger. This may relate to the transition conditions we adopt at $\rmag$.

Figure \ref{fig:sedlowmdot} shows the corresponding spectral energy distributions (SEDs) for these models. Because this accretion rate is fairly low (in Eddington unit, $\dot{M}/\medd$), there are three bumps in each spectrum, i.e. the synchrotron (peaks around $10^{11-12}$ Hz for the chosen parameters), the first-order inverse Compton scattering (peaks around $10^{14-16}$ Hz), and the bremsstrahlung (peaks around $\sim k T_{\rm e}\sim 10^{19-20}$ Hz). Emission from higher-orders Comptonization is weaker than the bremsstrahlung, and it is invisible here. One quick finding from this plot is that the SED of MAD is similar to that of SANE (see the MAD emission by \citealt*{Ryan2017, Ryan2018, Chael2019} and the SANE emission by \citealt*{Manmoto1997, Yuan2003}). This is not surprising, not only because they share the same radiative mechanisms, but also because the dense spirals are only weakly magnetized (cf.\ Figure \ref{fig:dynlowmdot}), similar to that of SANE (especially SANE with weak global fields). Emission of low-$\Phi_{\rm n}$ MAD is very close to that of near-critical SANE, only that the synchrotron bump of low-$\Phi_{\rm n}$ MAD is slightly lower in both the peak frequency and the peak luminosity. Both are a direct consequence of the decrease in magnetic field strength (increase of $\beta$) in the gas regions, see bottom-right panel of Figure \ref{fig:dynlowmdot}.

For the radiation of MAD, we additionally observe two results from Figure \ref{fig:sedlowmdot}. First, the Compton bump is relatively brighter (with respect to the synchrotron bump) in MAD systems, as optical depth increases with $\Phi_{\rm n}$. This phenomena should be more evident at higher accretion rate (or luminosities), since Compton scattering is sensitive to the optic al depth (or equivalently $\Sigma$, e.g., \citealt*{Rybicki1979, Dermer1991}). Secondly, we may observe a weak spectral offset at high-frequency radio bands, due to the sharp transition in density and magnetic fields (within the gas) at $\rmag$. In hot accretion flows, the emission site of synchrotron moves inward with increasing photon energy (e.g. \citealt*{Narayan1998}). If $\rmag<10$--$20\rg$, we may observe a weak offset/shift in radio bands, as shown by the blue dashed curve in Figure \ref{fig:sedlowmdot} (see the {\it right panel} for a zoom-in of the synchrotron bump, the offset happens at around $\nu_{\rmag} \approx15$--30 GHz). Note that the exact location of $\nu_{\rmag}$ depends on various parameters, among which $\rmag$, $\dot{M}/\medd$ and $\mbh$ play key roles. 

\begin{figure}
  \centering
  \includegraphics[width=8.cm]{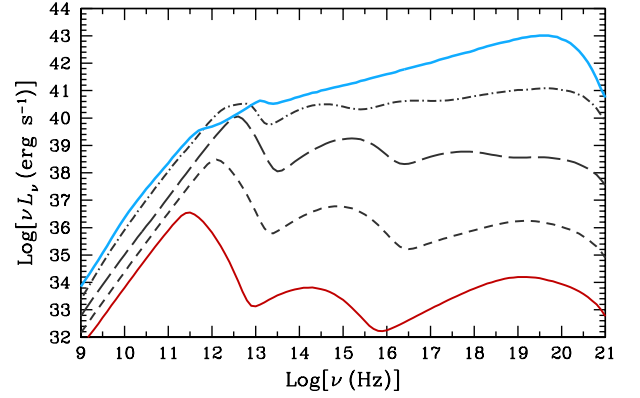}
  \caption{SED of high-$\Phi_{\rm n}$ ($\betazrout = 1$ with $R_{\rm out} = 10^3 \rg$) MAD systems at different accretion rate. From top to bottom, the boundary accretion rates at $R_{\rm out}$ are respectively, $\dot{M}_0 = 10^{-1}\medd$ (blue solid), $10^{-2}\medd$ (black dot-dashed), $10^{-3}\medd$ (black long dashed), $10^{-4}\medd$ (black dashed) and $10^{-5}\medd$ (red solid, the same to the red solid curve in Figure \ref{fig:sedlowmdot}).}\label{fig:sedvarmdot}
\end{figure}

\subsection{Radiative Properties at Different $\dot{M}/\medd$}\label{sec:varmdot}

Figure \ref{fig:sedvarmdot} shows the spectral properties of MAD with increasing accretion rate $\dot{M}_0/\medd$. We consider the high-$\Phi_{\rm n}$ MAD case, where we adopt $\betazrout = 1$ at $R_{\rm out} = 10^3 \rg$. All the systems enter the spiral/void region at the outer boundary, i.e. $\rmag > R_{\rm out}$. Other parameters except the accretion rate are the same to those in Section \ref{sec:lowmdot} (and see Table \ref{tab:par}). For curves from top to bottom in Figure \ref{fig:sedvarmdot}, the boundary accretion rates at $R_{\rm out}$ are respectively, $\dot{M}_0/\medd = 10^{-1}$ (blue solid curve), $10^{-2}$ (black dot-dashed), $10^{-3}$ (black long dashed) and $10^{-4}$ (black dashed). For reference, the bottom red solid curve shows the $\dot{M}_0/\medd = 10^{-5}$ case, which is the same to that in Figure \ref{fig:sedlowmdot}.

As shown in Figure \ref{fig:sedvarmdot}, the spectral evolution of MAD with increasing accretion rates is quite similar to that of SANE (cf. figure 1 in \citealt*{Yuan2014}). First, as $\dot{M}$ increases, the peak frequency of the synchrotron bump, which depends on both the electron temperature and magnetic field strength\citep{WZ2000}, moves to higher frequencies. Second, as $\dot{M}$ increases, the soft photon power $L_{\rm soft}$ increases much slower than the power supplied to the electrons $L_{\rm hard}$ \citep{Yang2015, Zdziarski2002}; higher-order Compton scattering dominants the emission, and the spectrum will be of power-law shape in X-rays at high accretion rate. The strong radiative cooling due to the Compton scattering processes will lead to a decrease in the electron temperature.

We note that such trend is universal, i.e. similar trend is always observed (not shown here) for models with other model parameters.

\subsection{Radiative Efficiency and Critical Accretion Rate} \label{sec:lcrit}

One of the most important quantities in accretion theory is the radiative efficiency $\eta$. Through radiative feedback, the efficiency $\eta$ also plays an important role in the coevolution of supermassive BH (SMBH) and its host galaxy over cosmic time (see e.g., \citealt*{Merloni2008, Mocz2013}). Meanwhile the 
SMBH mass function evolving over cosmic time can also be established based on both $\eta$ and the observed hard X-ray luminosity function. The radiative efficiency describes the efficiency in converting rest-mass energy into radiation, which can be defined as 
\begin{equation}
\eta = {L_{\rm bol}\over \dot{M}_{\rm BH} c^2}.\label{eq:eff}
\end{equation}
Here $L_{\rm bol}$ is the bolometric luminosity. For accretion with outflow ($\dot{M}$ is not a constant), we adopt the net accretion rate (onto BH) $\dot{M}_{\rm BH}$ as a representative \citep{Xie2012}. Depending on BH spin, the efficiency of SSD lies in the range $5.7-42\%$, with a typical value $10\%$. Hot accretion flow is known to be radiatively inefficient. Its radiative efficiency $\eta$ has a positive but complex dependence on mass accretion rate \citep{Narayan1998, Xie2012}.

Hot accretion flows can only exist below a certain critical accretion rate ($\dot{M}_{\rm crit, MAD}$ and $\dot{M}_{\rm crit, SANE}$ respectively, for MADs and SANEs), above which radiative cooling exceeds the heating. In this case, the accretion flow will become two-phase, i.e. numerous small cold and dense clumps are formed, embedded in and coupled with the hot but tenuous gas (e.g. \citealt*{YF2003, Wu2016}). At even higher accretion rates, cold clumps may merge and settle down to the midplane, leading to the disk/SSD-corona configuration \citep{Yang2015, Wu2016}.

We focus on the purely-hot phase accretion case, i.e. the clumpy two-phase accretion with global magnetic fields will not be investigated here. In addition, the effects of $\alpha$ and $\delta$ on the critical accretion rate as well as the radiative efficiency, as investigated by \citet*{Xie2012}, will also not be probed here. Instead, we focus on the effect of global magnetic fields, through parameter $\betazrout$. To simplify numerical calculations, we further set the boundary to $R_{\rm out} = 200 \rg$. For this choice of $R_{\rm out}$ and outflow parameter $s$, the net accretion rate onto BH will be $\dot{M}_{\rm BH} \approx 0.25 \dot{M}_0$.

\begin{figure}
  \centering
  \includegraphics[width=8.5cm]{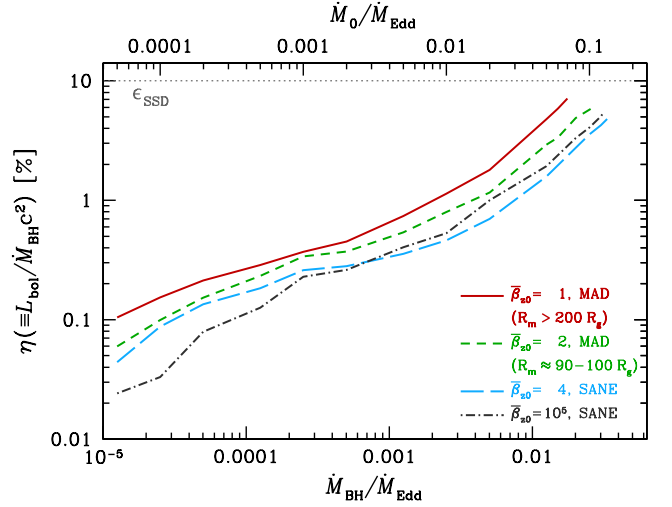}
  \caption{Radiative efficiency of MAD and SANE systems. The outer boundary is set to $R_{\rm out} = 200 \rg$, and other basic parameters are the same to those listed in Table \ref{tab:par}. As labelled in the figure, we show results for $\betazrout = 1$ (red solid, MAD with $\rmag > R_{\rm out}$), $\betazrout = 2$ (green dashed, MAD with $\rmag \approx 90$-$100 \rg$). For comparison, we also show results for two type of SANEs, one with $\betazrout = 4$ (blue long-dashed, near-critical SANE) and the other with $\betazrout=1\times10^5$ (black dot-dashed, standard SANE). The dotted curve shows the efficiency of SSD, i.e. $\eta=10\%$. }\label{fig:radeff}
\end{figure}

The results are shown in Figure \ref{fig:radeff}. The red solid curve is for high-$\Phi_{\rm n}$ MAD with $\rmag > R_{\rm out}$ (we set $\betazrout=1$), and the green dashed curve is for MAD with $\rmag \approx 90$-$100\rg$ ($\betazrout=2$, the exact value of $\rmag$ weakly depends on accretion rate). We observe an enhancement in radiative efficiency for the increase in the strength of global magnetic fields, i.e. for a given $\dot{M}$ and outflow parameter $s$, the radiative efficiency is about 20-50\% higher in $\betazrout=1$ systems than in $\betazrout=2$ systems. Such enhancement is even higher at larger $\dot{M}$, i.e. when $\dot{M}_{\rm BH}>3\times10^{-3}\medd$, it will be larger by $\sim50\%$. Because of the difference in the radiative cooling, the critical accretion rate also differs, weaker cooling (lower $\betazrout$) results higher critical accretion rate (note that $\alpha$ is pre-fixed in all our calculations). For example, the critical accretion rate onto BH is $\dot{M}_{\rm BH,crit,MAD}\approx 1.8\times10^{-2}\medd$ for $\betazrout=1$, while $\dot{M}_{\rm BH,crit,MAD}\approx 2.7\times10^{-2}\medd$ for $\betazrout=2$. The maximal efficiency in both cases is $\eta\approx$ 7-8\%.

For comparison, we also shown in Figure \ref{fig:radeff} the blue long-dashed curve is for near-critical SANE with $\betazrout=4$, and the black dot-dashed is for standard SANE with $\betazrout=10^5$. For the SANE systems, we find that the radiative efficiency increases with increasing $\betazrout$ when synchrotron emission dominants (i.e. $\dot{M}_{\rm BH} < 4\times10^{-4}\medd$). Above this accretion rate, the efficiency is almost independent of the $\betazrout$, with difference less than 10\%. This is totally different to the change in the turbulent magnetic field strength, where stronger turbulent fields typically lead to an enhancement in radiation (e.g., \citealt*{Manmoto1997, Xie2010}). We find that in SANE systems, the effects of global fields are partially compensated by the gas pressure and turbulent stress terms, i.e. systems with stronger global fields (thus lower $\beta$) will have lower electron temperature, leading to weaker Compton scattering. Obviously, their SED will be dramatically different, as demonstrated in the subsequent section. The maximal efficiency of SANE is $\approx$6\%.

Our results also show that the efficiency depends on the mass accretion rate, i.e. $\eta(\mdot_{\rm BH}) \propto \mdot_{\rm BH}^{0.55-0.7}$, and the correlation slope becomes steeper at higher $\mdot_{\rm BH}$ (see also \citealt*{Xie2012} for the standard SANE case). There is no clear difference in the slope between MAD and SANE, especially at high $\mdot_{\rm BH}$ regime. Note that the correlation slope is systematically shallower than that of the $\eta(\mdot_{\rm BH}) \propto \mdot_{\rm BH}$ correlation, which is widely adopted in literature (e.g., \citealt*{Narayan1998, Merloni2008}, most of which are based on conventional version of hot accretion model that omit electron viscous heating by adopting $\delta \ll 1$.), i.e. low-$\mdot$ systems are actually brighter than previously thought.

For given $\alpha$ and other model parameters, one direct consequence of the differences in $\eta$ and $\dot{M}_{\rm crit}$ is that MAD shares a similar maximal luminosity to SANE. However, we caution that this result depends on the microphysics of turbulent heating, where we adopt the same prescription for the turbulent heating driven by MTI (MAD case) or MRI (SANE case) processes, cf. Sec. \ref{sec:maddyn}.

\begin{figure}
  \centering
  \includegraphics[width=8.cm]{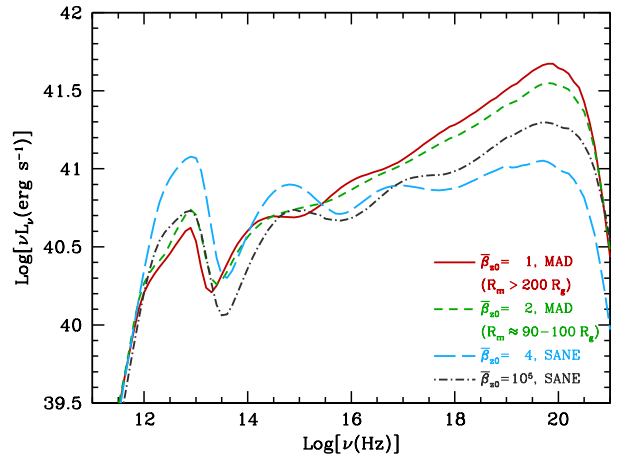}
  \caption{SED of MAD and SANE systems under the same accretion rate ($\dot{M}_0 = 1\times10^{-2}\medd$ at $R_{\rm out} = 200 \rg$). As labelled in this figure, the meaning of each curve is the same to that in Figure \ref{fig:radeff}.}\label{fig:sed_mad_sane}
\end{figure}

\subsection{Comparison between SANE and MAD}\label{sec:sane.vs.mad}

One notable fact of MAD system is that, although it is supplied with large amount of magnetic flux, the gas spirals actually have relatively weaker magnetic fields, with $\beta$ similar to SANEs of moderate magnetic flux (see also \citetalias{McKinney2012}). Beside, its radiative processes and the consequent spectrum are the same to those of SANE. All these make it challenging to distinguish MAD and SANE based on spectral information only. On the other hand, such disadvantage also implies that (at least some) low-luminosity AGNs and BHBs in their hard and intermediate states, whose central engine is argued to be SANE (for reviews, see \citealt*{Done2007, Ho2008, Yuan2014}), may possibly be MAD instead.

The spectral difference between MADs and SANEs under the same boundary accretion rate $\dot{M}_0$ is illustrated already in Figure \ref{fig:sedlowmdot} when the accretion rate is low. We here show the results at higher accretion rates. For this purpose, we take data from Section \ref{sec:lcrit}, where the boundary accretion rate at $R_{\rm out} = 200 \rg$ is chosen to be $\dot{M}_0 = 1\times10^{-2}\medd$ (thus $\dot{M}_{\rm BH} \approx 2.5\times 10^{-3}\medd$). The SEDs are shown in Figure \ref{fig:sed_mad_sane}, where each curve is of the same meaning to that in Figure \ref{fig:radeff}. We find that even with similar radiative efficiencies, the two SANEs show clear differences, i.e. the near-critical SANE (with $\betazrout=4$) is brighter in its synchrotron bump (because its magnetic field strength in gas regions is stronger), while the standard SANE (with $\betazrout=10^5$) is brighter in its Compton scattering bumps (mostly because the electrons are hotter). 

The two MADs are systematically brighter than the two SANEs, as already known based on the radiative efficiencies. When compared to SANEs, we notice that the synchrotron emission of MAD systems is actually weaker, but their Compton upper-scattered emission is much brighter. The first is due to weaker magnetic field strength in gas regions, while the latter is mainly due to higher optical depth, see Figure \ref{fig:dynlowmdot}. On average, under the same accretion rate, the X-ray luminosity of MAD is systematically brighter by a factor of 3-5 than that of SANE.

We note that for given luminosities, MAD inclines to have higher surface density but lower electron temperature. Such information may be revealed from detailed modeling of the Comptonized spectrum, e.g., in hard X-ray band. We will discuss this point later in Section\ \ref{sec:bhb}.

\section{Discussions}\label{sec:discussion}

\subsection{Accretion Mode in Sgr A* and M87} \label{sec:gcm87}

Extensive efforts have been devoted to investigate the accretion mode in the two nearby low-luminosity AGNs with SMBHs, i.e. Sgr A* and M87 (e.g., \citetalias{Narayan2003, McKinney2012}; \citealt*{Yuan2003, Shcherbakov2013, Gold2017, Ryan2017, Ressler2017, Ressler2019, EHT2019, Chael2019}), where the innermost $R<10^{1-2}\rg$ regions can be resolved by current facilities such as the Event Horizon Telescope (EHT) and the GRAVITY-VLT. However, although both sources have numerous multi-band (and multi-epoch) observations, the magnetic flux near BH still remains uncertain. It is also clear from this work that the SED modelling itself is very difficult to distinguish between SANE and MAD (see also \citealt*{Gold2017, Ressler2017, EHT2019}).

Based on linear polarization information, \citet{Gold2017} argue that Sgr A* favors the MAD interpretation. However, evidence in favor of SANE also exist. Sgr A* is fed by stellar wind from a cluster of O and Wolf-Rayet stars orbiting around the BH \citep{Cuadra2008, Wang2013}. Recently \citet*{Ressler2019} carried out MHD simulations to model these gases, and found the magnetic flux around Sgr A* is only about 10\% the minimal of MAD. Besides, an infrared ``hot spot'' flare with Keplerian rotation is detected by the high-resolution GRAVITY-VLT Interferometer \citep{Gravity2018}, where a strong poloidal magnetic field is suggested by polarization signature of the spot. Although high-$\Phi_{\rm n}$ MAD model can provide strong poloidal field, the rotational of dense spiral/blob should be highly sub-Keplerian. We note that the GRAVITY-VLT observation is still consistent with a low-$\Phi_{\rm n}$ MAD interpretation, in which case we may detect a weak spectral shift/offset around 10-300 GHz, cf. the green long-dashed curve in Figure \ref{fig:sedlowmdot} and note the difference in BH mass. 
%It is worthwhile to carry out detailed modeling of Sgr A* under different accretion models. 
%For Sgr A*, there are possibly two additional ways to discriminate the two models, both rely on future EHT observations at even higher frequencies. One is through the polarization in the visibility (Fourier) domain and the other is through timing properties \citep{Gold2017}, both are out of the scope of this work.

M87 is a nearby AGN suggested to be in MAD \citepalias{Narayan2003, Narayan2012, McKinney2012}. It has a powerful jet extending up to kpc scales and two giant ``X-ray cavities'' \citep{Allen2006}. The X-ray cavity may be created by the deposit of the jet power into the ambient gas within the galaxy. The jet power is found to be comparable or even higher than the bolometric luminosity in M87 (\citealt*{Allen2006}, see \citealt*{EHT2019} for a summary). With a combination of jet power, the EHT imaging at 230 GHz and the constraints in X-ray luminosity, M87 favors a MAD interpretation, although SANE models around maximal spinning BH cannot be ruled out \citep{EHT2019}. Based on two-temperature MAD simulations, \citet*{Chael2019} find that MAD model can produce not only the broadband SED but also the core-shift effects observed in high-resolution radio images \citep{Hada2011}. 

\subsection{AGNs}

AGNs are known to have a large diversity (spanning 3-4 orders of magnitude) in their radio-loudness, which is defined as the ratio of jet-traced radio luminosity to accretion disk-traced optical-UV luminosity. Most investigations on MAD-type accretion in AGNs focus on radio-loud (RL hereafter) sources, including RL galaxies and blazars. 
%Theoretically the power of a BZ-type jet is determined by magnetic flux threading the BH \citep{Sikora2013} and also by BH spin. In this spirit, RL and radio-quiet (RQ) sources may be in MAD and SANE states, respectively. Indeed, some 
Some RL AGNs have jet efficiencies (defined as the ratio of jet power to accretion power) greater than $100\%$ \citep{Rawlings1991, Ghisellini2010, Ghisellini2014, Fernandes2011, McNamara2011}. Similar conclusion can also be reached on a statistical basis. For example, for a sample of blazars, the jet power is found to be greater than, but tightly correlates with, the accretion power \citet*{Ghisellini2014}. Another example comes from \citet*{Z2014} (updated later by \citealt*{Zdziarski2015}), where they reported a tight correlation between the jet magnetic field flux and disk luminosity based on a sample of 76 RL active galaxies. Such correlation is exactly what MAD predicts. All these observations require the extraction of BH spin power, or equivalently, they are in favor of the MAD model.

Recently, \citet*{Grupta2018, Grupta2019} investigated the X-ray properties of luminous RL galaxies and their radio-quiet (RQ hereafter) counterparts, based on a {\it Swift}/BAT sample with similar black hole mass and Eddington ratio. They use mid-infrared emission as a proxy of cold disk emission, and found that RL AGNs are X-ray-louder than RQ AGNs, but their X-ray spectral slope is similar. \citet*{Grupta2019} interpreted this result as an effect of the difference in BH spin. In our understanding, it is also consistent with the ``RL in MAD and RQ in SANE'' interpretation. Under the same $\mdot$ MAD is brighter by only $\sim$ 2 than SANE. Meanwhile, the optical depth is higher in MAD (compared to SANE), thus it will be brighter in hard X-rays, see Figure \ref{fig:sed_mad_sane}.

Finally we caution that radio-moderate or even RQ systems can still be in MAD, i.e. a maximal-power jet does not guarantee luminous in jet emission. The energy-to-emission conversion could be highly suppressed in jet for several reasons \citep{Gold2017}, e.g., too few electrons due to inefficient mass-loading, and/or low energy dissipation power to accelerate electrons to relativistic energies. 

%Another issue is that a significant fraction of RL sources in fore-mentioned investigations belong to bright AGNs, whose central engine is possibly cold accretion. As a contrast, the hot MAD explored here applies to low-luminosity AGNs \citep{Ho2008, Yuan2014}, although it may be generalized to describe the corona of bright AGNs.

\subsection{Black hole X-ray binaries} \label{sec:bhb}

Thanks to relatively short evolution time, BHBs reveal fruitful information on the underlying accretion physics. The leading picture for hard and hard-intermediate states (see \citealt*{Belloni2010} for state classification) is the accretion-jet model \citep{Zdziarski2004, Done2007, Yuan2014}, where the cold disk truncated at a radius $R_{\rm tr}$ and a hot flow inside that radius \citep{Esin1997, YCN2005}. In this model, the hard X-rays originate from the hot accretion flow. 

Whether the hot accretion flow is MAD or SANE remains unclear. One observational feature of BHBs is the hysteresis behavior, i.e. the hard-to-soft state transition happens at a brighter luminosity ($\sim 10\% \ledd$) compared to that of the soft-to-hard transition ($\sim 1\% \ledd$). \citet*{Begelman2014} considered the effect of advection and accumulation of the magnetic flux generated at $R_{\rm tr}$. They found that the accretion becomes highly (weakly) magnetized during the rising (decline) part of the outburst. The effective $\alpha$ of hot accretion flow will be different in the two parts. This model provides a plausible interpretation for the hysteresis. However, the possibility of entering the MAD phase is not mentioned in \citet*{Begelman2014}. Moreover, as discussed in Section \ref{sec:lcrit}, we find that there is no notable difference in critical luminosity between MAD and SANE, unless the effective $\alpha$ (suggested in \citealt*{Begelman2014}) and/or the dissipative heating are revised. The SANE with stronger turbulent magnetic fields is still possible for the rising part.

Broad-band ($\sim$1--100 keV) X-ray spectra in hard state show high-energy cut-offs at $\sim$50--200 keV (in $\nu L_\nu$ plots, e.g., \citealt{Zdziarski1998, Done2007, Joinet2008, Miyakawa2008, Yan2020} and references therein). Together with the photon index, we can constrain both the electron temperature $T_{\rm e}$ and the optical depth $\tau$. We expect to observe a jump (from time evolution and its correlation with X-ray luminosity $L_{\rm X}$) in optical depth when the system transits from SANE into MAD during the rising phase. Besides, with a comparison of the normalization and/or the slope of correlations of $\tau$-$L_{\rm X}$ and $T_{\rm e}$-$L_{\rm X}$, we may quantitatively examine the ``MAD in rising and SANE in decline'' hypothesis in future.

%these quantities between the rising part and the decline part, 

\section{Summary} \label{sec:summary}

MAD is achieved when the accretion system is fed with sufficient amount of magnetic flux \citepalias{Narayan2003}. It provides an efficient way to launch a spin-powered BZ-jet. Observational evidence on MAD in RL systems is now accumulating (e.g., \citealt*{Ghisellini2014, Z2014, Zdziarski2015}). However, its application is limited by the lack of spectral calculations. The only existing calculations focus on dimmest ones \citep{Gold2017, Chael2019}, which cannot be applied to systems at higher $\mdot/\medd$. 

%awaits detailed spectral calculations, especially at higher $\mdot$.

In this work we investigate the radiative properties of MAD systems. We construct the global magnetic fields based on 3D MHD simulations (e.g., \citetalias{McKinney2012, Narayan2012}), and develop height-integrated equations to describe the dynamics of MAD. The transition from the outer axisymmetric region to the inner spiral/void region is determined automatically. One advantage for this treatment is that our model can describe both MAD and SANE. Consistent with MAD simulations \citepalias{Igu2008, McKinney2012}, we find a significant fraction of magnetic field is expelled from dense gas spirals into magnetized voids, due to MII process. The spirals are weakly magnetized ($\beta_{\rm s}\ga$1), while the dilute voids have $\beta_{\rm v}\ll$1.

The radiative processes in MAD systems are synchrotron, bremsstrahlung and the Compton scattering, the same to those of SANEs. We find that the MAD shares a similar spectrum to that of SANE with global fields. Although this result makes it challenging to discriminate MAD from SANE, it provides a natural explanation on the spectral similarities between RL and RQ AGNs, if RL ones are in MAD while RQ ones are in SANE (\citealt*{Sikora2013}, but see \citealt*{Grupta2019} for the effect of BH spin.)

We further investigate the radiative efficiency of MAD, and find that for a given accretion rate MAD is systemically brighter than SANE. MAD also have a higher maximal radiative efficiency than SANE, with the former 7--8\% and the latter about $6\%$. On the other hand, MAD has a lower (by a factor of $\sim$1.5--2) critical accretion rate than SANE. These two results imply that for given other parameters, the maximal luminosity of MAD will actually be comparable (but slightly lower) than that of SANE. We note that the maximal luminosity of MAD can be higher, if it has a higher viscosity $\alpha$ and/or a more efficient energy dissipation ($q_{\rm heat}$).

\section*{Acknowledgments}
We appreciate Marek Sikora, Feng Yuan, Shuang-Liang Li, Zhaoming Gan and De-Fu Bu for helpful discussions, and Can Cui for the guiding plots of magnetic fields of numerical simulations of MAD systems. This work was supported in part by the National Program on Key R\&D Project of China (Grants 2016YFA0400804), the Natural Science Foundation of China (grants 11873074, 11573051, 11633006, and 11661161012), the K.C. Wong Education Foundation of CAS, and the Key Research Program of Frontier Sciences of CAS (No. QYZDJ-SSW-SYS008). F.G.X. was also supported in part by the Youth Innovation Promotion Association of CAS (id. 2016243) and the Natural Science Foundation of Shanghai (grant 17ZR1435800). A.A.Z. was supported in part by the Polish National Science Centre grant 2015/18/A/ST9/00746.

\begin{appendix}

\section{Key Assumptions and Height-Integrated Dynamical Equations for MAD Systems}\label{app:madeqs}

Below we provide the details on the derivation of the dynamical equations listed in Section \ref{sec:maddyn}. Revelant information can also be found in e.g., \citealt*{Manmoto1997, LC2009, Cao2011, Oda2012}. We adopt a cylindrical coordinate ($R, \phi, z$) and take the pseudo-Newtonian potential (\citealt{PW1980}. The PW potential $\psi$ can be expanded as a sum of series of powers of $(z/R)$ \citep{Manmoto1997}, i.e. 
\begin{equation}
\psi(R,z) = -{GM_{\rm BH}\over R-2\rg}+{1\over 2}\Omega_{\rm K}^2 R^2\left({z\over R}\right)^2+O\left(\left({z\over R}\right)^4\right).
\end{equation}
Here $\Omega_{\rm K}(R)$ (defined as $\Omega_{\rm K}^2(R) = = {G \mbh R^{-1}(R-2 R_{\rm g})^{-2}}$) is the Keplerian angular velocity at the mid-plane. We thus have,
\begin{equation}
{\partial\psi\over\partial R} \approx \Omega_{\rm K}^2 R \!\left[1+{{\rm d}\ln\Omega_{\rm K}\over {\rm d}\ln R} \left({z\over R}\right)^2\right], \hspace{0.5cm}
{\partial\psi\over\partial z} \approx \Omega_{\rm K}^2R \!\left({z\over R}\right).
\end{equation}

The magnetic fields in MAD system are seperated into two types. One is the tangled turbulent one, which is determined through the plasma parameter $\beta_{\rm t} = P_{\rm g}/P_{\rm tm}$. The other is the ordered global one, which we follow the results from 3D MAD simulations \citepalias{Narayan2012, McKinney2012}. We first assume all the field components have no azimuthal dependence within either the spirals or voids; however, the field strengths are different between them. We then consider their distribution in the vertical ($z$) direction. $B_z(R,z)$ is assumed to be a constant along $z$, while the radial ($B_R(R,z)$) and the azimuthal ($B_\phi(R,z)$) components are assumed to be odd and linear functions of $z$, i.e. $B_R(R,z) = \br\, (z/H)$ and $B_\phi(R,z) = \bphi\, (z/H)$. Here $\br$ and $\bphi$ are the magnetic field strengths at the surface of the accretion flow (i.e., defined at $z=H$. Note that in our approximation $B_R(R,0) = B_\phi(R,0) = 0$), while $\bz$ is that at the mid-plane of the flow. Without losing generality, we further assume the accretion flow rotates with $\Omega > 0$ and take $\bz > 0$. Consequently we have $\br > 0$ and $\bphi < 0$ (note that $\br\bphi < 0$). In other words, we have $\pr > 0$ and $\kappa_\phi < 0$, as shown in Table \ref{tab:par}. We then consider their radial distribution. We assume the fields roughly follow the $R^{\sim -1}$ profile in the radial direction \citepalias{McKinney2012}. We thus have,
\begin{equation}
{\partial B_R(R, z)\over\partial R}\approx - {\br\over R}, \hspace{0.5cm}{\partial B_R(R, z)\over \partial z} \approx {\br\over H}, \hspace{0.5cm}{\partial B_\phi(R, z)\over \partial R}\approx -{\bphi\over R}, \hspace{0.5cm}{\partial B_\phi(R, z)\over\partial z} \approx {\bphi\over H}, \hspace{0.5cm}{\partial B_z(R, z)\over\partial z} = 0.
\end{equation}
Considering the vertical distribution, we approximate the vertically-averaged magnetic field strength within the accretion flow at a given radius $R$ as,
\begin{equation}
\overline{B_R(R,z)} = \br/2, \hspace{0.5cm}\overline{B_\phi(R,z)} = \bphi/2, \hspace{0.5cm}\overline{B_z(R,z)} = \bz.
\end{equation}

We now consider the dynamical equations to describe the MAD system. 
We assume the accretion flow is in steady state ($\partial/\partial t=0$), which is adequate for applications of time-averaged properties. We consider an accretion flow with scale height $H$, above which outflow exists. We further assume the accretion flow is quasi-static in the vertical direction, i.e. $V_z\approx 0$ within the accretion flow. 

The first is the mass conversation equation of the of accretion flow, which can be described as
\begin{equation}
{\partial\over\partial R}\left(R\rho V_R\right) = -R{\partial\over\partial z}\left(\rho V_z\right). \label{eq:contapp2}
\end{equation}

Under a height-integration, the left hand side of Equation \ref{eq:contapp2} represents the change in accretion rate, while the right hand side represents the mass loss rate. Following \citet{BB99}, we replace the height-itegrated form of Equation (\ref{eq:contapp2}) with the power-law profile of mass accretion rate, as shown in Equation (\ref{eq:mdot}).

The second equation is the dynamical equation, which reads
\begin{eqnarray}
\rho (\vvect\cdot)\vvect & = & -\nabla (P_{\rm g} + P_{\rm tm}) - \rho \nabla\psi + {1\over4\pi}(\nabla\times\bvect)\times\bvect + \nabla\cdot{\bf T} \nonumber \\
& = &  -\nabla (P_{\rm g} + P_{\rm tm} + P_{\rm m}) - \rho \nabla\psi + {1\over4\pi}(\bvect\cdot\nabla)\bvect + \nabla\cdot{\bf T}. \label{eq:dynapp1}
\end{eqnarray}
In the second expression we have taken the identity ${1\over4\pi}(\nabla\times\bvect)\times\bvect = -\nabla(B^2/8\pi) + {1\over4\pi}(\bvect\cdot\nabla)\bvect$. The total pressure is defined as $P_{\rm tot} = P_{\rm g} + P_{\rm tm} + P_{\rm m}$, where $P_{\rm m} = B^2/8\pi$. In MAD systems, the MHD turbulence is driven by MII \citep{White2019}, while in SANE the MHD turbulence is driven by MRI \citep{Balbus1998}. In this work, the stress tensor due to both hydro and MHD turbulence is described by ${\bf T}$, where the only non-zero components considered are the $\phi r$ term $\tau_{\phi r}$ and the symmetric $r\phi$ one $\tau_{R\phi}$. We have $\tau_{\phi R} = \tau_{R\phi}$. Consequently, the non-zero stress force term is in the azimuthal direction, which takes the form $(\nabla\cdot {\bf T})_\phi ={1\over R^2}{{\rm d}\over {\rm d}R}(R^2\tau_{\phi R})$. We adopt the conventional $\alpha$-description \citep{SS1973} that $\tau_{\phi R}\approx -\alpha (P_{\rm g} + P_{\rm tm})$.

We first consider the vertical direction part of Equation (\ref{eq:dynapp1}). Because we assume the accretion flow is in the hydro-static condition in $z$ direction, it can be re-written as
\begin{eqnarray}
{\partial P_{\rm tot}\over\partial z} & = & -\rho {\partial\psi(R,z)\over\partial z}  + {1\over 4\pi} B_R {\partial B_z(R,z)\over\partial z} \nonumber\\
& \approx & - \rho \Omega_{\rm K}^2\ z. \label{eq:zapp1}
\end{eqnarray}
For the second expression above, the magnetic tension term can be omitted (but see, e.g., \citealt*{Cao2011}). In the above expression, we adopt the approximation ${\partial\psi(R,z)\over\partial z} \approx -\Omega_{\rm K}^2\ z$, which is valid when $H/R\ll 1$ (see \citealt*{Gu2007} for a more careful treatment). For an isothermal accretion flow, $\partial P_{\rm tot}/\partial z= c_{\rm s}^2 \partial\rho/\partial z$, where isothermal sound speed $c_{\rm s}$ is defined through $c_{\rm s}^2= P_{\rm tot}/\rho$. The solution to Equation\ (\ref{eq:zapp1}) provides a density profile in vertical direction, i.e. 
\begin{equation}
\rho(R, z) = \rho(R, 0)\exp\left(-{z^2\over 2} {\Omega_{\rm K}^2\over c_{\rm s}^2}\right)\equiv \rho(R, 0)\exp\left(-{z^2\over 2H^2}\right),\label{eq:densityz}
\end{equation}
where the scale-height $H$ is determined by $H=c_{\rm s}/\Omega_{\rm K}$.

The height-integrated dynamical equation in radial direction can be read as
\begin{eqnarray}
V_R{{\rm d} V_R\over {\rm d}R} + (\Omega_{\rm K}^2 - \Omega^2)\ R+ {1\over\rho} {dP_{\rm tot}\over dR} & = & {1\over 4\pi} {\int \left[B_R {\partial B_R\over\partial R} + B_z {\partial B_R\over\partial z} - {B_\phi^2\over R}\right] {\rm d}z \over \int \rho dz} \nonumber\\
& \approx & {1\over 4\pi \rho }\left({\br\bz\over H} - {\br^2+\bphi^2\over 4R}\right) \nonumber\\
& \approx & {1\over \rho H} {\br\bz\over 4\pi}. \label{eq:rapp2}
\end{eqnarray}

For the dynamical equation in azimuthal direction, before height integration it reads,
\begin{equation}
{\rho V_R\over R} {{\rm d}\over {\rm d}R}\left(\Omega R^2\right) = {1\over R^2} {{\rm d}\over {\rm d}R}\left(R^2 \tau_{\phi R} \right) + {1\over 4\pi} \left[B_R {\partial B_\phi\over \partial R} + B_z {\partial B_\phi\over \partial z} + {B_\phi B_R\over R}\right].\label{eq:phiapp3}
\end{equation}
For the height-integration of the global field parts, we adopt the following expressions, i.e.
\begin{eqnarray}
\int {\rm d}z \left[B_R(R,z) {\partial B_\phi(R,z)\over \partial R} + {B_R(R,z) B_\phi(R,z)\over R}\right] & \approx & \int {\rm d}z \left[-{B_R(R,z) B_\phi(R,z)\over R} + {B_R(R,z) B_\phi(R,z)\over R}\right] = 0, \nonumber\\
\int {\rm d}z \left[B_z(R,z) {\partial B_\phi(R,z)\over \partial z}\right] & = & \bz \int {\rm d}z {\partial B_\phi(R,z)\over\partial z}\approx \bz\bphi. \nonumber
\end{eqnarray}

Under the assumptions that in the vertical direction the accretion flow is isothermal and $\Omega$ is a constant, the height-integrated form of Equation (\ref{eq:phiapp3}) will be,
\begin{equation}
-{1\over 4\pi}{{\rm d}\over {\rm d}R}\left({\mdot \Omega R^2\over C_{\rm m}}\right)\approx {{\rm d}\over {\rm d}R}\left(R^2 H\tau_{\phi R}\right) + R^2{\bz\bphi\over 4\pi}.\label{eq:phiapp4}
\end{equation}
Note that since $\br\bphi<0$ and $\bz\bphi<0$, the gloal field helps to transport angular momentum outward. In order to be consistent to SANE models (cf. \citealt*{Yuan2014}), we additionally include the eginvalue $j_0$ in Equation (\ref{eq:phiapp4}). The angular momentum equation then reads as Equation (\ref{eq:omegadyn}).
%\begin{equation}
%-{1\over 4\pi C_{\rm m}}{d\over dR}\left[\mdot(\Omega R^2-j_0)\right] \approx {d\over dR}\left(R^2 H\tau_{\phi r}\right) + RH{1\over 4\pi} {\br\bphi\over 4}.
%\end{equation}
%This is the angular momentum equation adopted in this work.

The final two equations describe the energy balance for ions and electrons, which read
\begin{eqnarray}
\nabla\cdot\left[(\rho\epsilon_{\rm i}+P_{\rm i})\vvect\right] - (\vvect\cdot\nabla)P_{\rm i} & \equiv & \rho \vvect\cdot(\nabla\epsilon_{\rm i} - {P_{\rm i}\over\rho^2} \nabla\rho) = (1-\delta) q_{\rm heat} - q_{\rm ie},\label{eq:ionapp1}\\
\nabla\cdot[(\rho\epsilon_{\rm e}+P_{\rm e})\vvect] - (\vvect\cdot\nabla)P_{\rm e} & \equiv & \rho \vvect\cdot(\nabla\epsilon_{\rm e} - {P_{\rm e}\over\rho^2} \nabla\rho) = \delta q_{\rm heat} + q_{\rm ie} - q_{\rm rad}. \label{eq:eapp1}
\end{eqnarray}
For the identities in the above expressions, we have taken advantage of $\nabla\cdot(\rho\vvect)=0$ (mass conservation. Or equivalently $\rho \nabla\cdot\vvect = -\vvect\cdot\nabla\rho$) to re-write in the forms of $\nabla\cdot(\rho\epsilon\vvect) = \epsilon\nabla\cdot(\rho\vvect) + \rho \vvect\cdot\nabla\epsilon = \rho \vvect\cdot\nabla\epsilon$ and $\nabla\cdot(P\vvect)-(\vvect\cdot\nabla)P = P(\nabla\cdot\vvect) = -{P\over\rho} \vvect\cdot\nabla\rho$. We then integrate Equations (\ref{eq:ionapp1}) and (\ref{eq:eapp1}) in the $z$-direction to derive the height-integrated energy equations, which are listed in Sec. \ref{sec:maddyn} as Equations (\ref{eq:ionapp}) and (\ref{eq:eapp}), respectively. 

\section{$C_{\rm m, min}$: the Minimal Covering Factor of Gas Spirals}\label{app:cmin}

\begin{figure}
\centering
\includegraphics[width=7.cm]{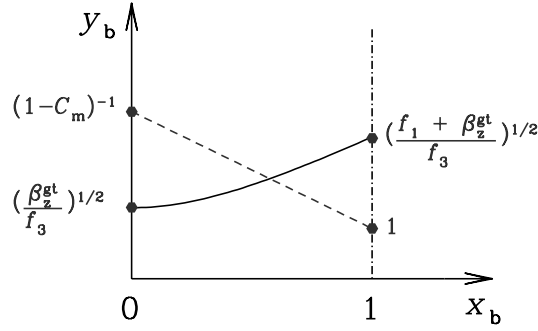}
\caption{Schematic plot for the $C_{\rm m,min}$ evaluation. We define $x_b = B_{z{\rm,s}}/\bar{B}_z$, $y_b = B_{z{\rm,v}}/\bar{B}_z$. The dot-dashed line marks the boundary at $x_b = 1$. The solid and dashed curves are, respectively, the hyperbolic equation (Equation\ (\ref{eq:ac1}), pressure balance) and the linear one (Equation\ (\ref{eq:ac2})). Their locations in $y$ axis at the two boundaries ($x_b = 0$ and $x_b = 1$) are marked by hexagons and are labeled with their corresponding $y_b$ values.}
\label{fig:cmin}
\end{figure}

The coupling between dense spiral and magnetized void is approximated by pressure balance between each other. Consequently, The covering factor of dense spirals, $C_{\rm m}$, has a lower-limit constraint, which is defined as the critical minimal value $C_{\rm m,min}$. Physically we should always have $C_{\rm m} > C_{\rm m,min}$. 

To evaluate the value of critical minimal value $C_{\rm m,min}$, we first define $x_b = B_{z{\rm,s}}/\bar{B}_z$, $y_b = B_{z{\rm,v}}/\bar{B}_z$, $f_{1} = 1 + ({B_{R{\rm,s}}^{s}}^2 + {B_{\phi{\rm,s}}^{\rm s}}^2)/(4 B_{z{\rm,s}}^2)$, $f_3 = 1 + {B_{\phi{\rm,v}}^{\rm s}}^2/(4B_{z{\rm,v}}^2)$ and $\beta_z^{\rm gt} = 8\pi (P_{\rm g,s}+P_{\rm tm,s})/\bar{B}_z^2 = 8\pi P_{\rm g,s}/\bar{B}_z^2 \times (1+\beta_{\rm t})/\beta_{\rm t}$. Then the hyperbolic Equation\ (\ref{eq:equalp}) and the linear Equation\ (\ref{eq:bav}) can be re-written as,
\begin{eqnarray}
{\rm hyper:}\hspace{0.5cm} & - f_{1}\, x_b^2 + f_3\, y_b^2&= \beta_z^{\rm gt}, \label{eq:ac1}\\
{\rm linear:}\hspace{0.5cm}& C_{\rm m}\, x_b + (1-C_{\rm m})\, y_b &=  1. \label{eq:ac2}
\end{eqnarray}

In order to have a clear illustration, we show the two equations in Figure\ \ref{fig:cmin}, i.e. solid curve for the hyperbolic one, and dashed curve for the linear one. Obviously since $f_1>f_3>1$, we should have $0 < x_b < 1 < y_b$. We emphasize that $y_b$ has a positive relationship with $x_b$ in Equation (\ref{eq:ac1}), while it has a negative relation in Equation (\ref{eq:ac2}). Consequently, there will be at most only one solution in the range $0 < x_b < 1$.

Since at one boundary $x_b=1$ we have $y_b {\rm (hyper)} = \left({f_1+\beta_{\rm z}^{\rm gt}\over f_3}\right)^{1/2} > y_b {\rm (linear)} = 1$, in order to have a pressure-balanced solution with $0 < x_b < 1$, we thus require that at the other boundary $x_b=0$ that
\begin{equation}
y_b {\rm (hyper)} = \left({\beta_z^{\rm gt} \over f_3}\right)^{1/2} < y_b {\rm (linear)} = {1\over(1-C_{\rm m})},
\end{equation}
which leads to the minimal value of $C_{\rm m,min}$,
\begin{equation}
C_{\rm m,min} = 1-\left({\beta_z^{\rm gt} \over f_3}\right)^{-1/2}\equiv 1- \left({1+\beta_{\rm t}\over f_3\beta_{\rm t}} {P_{\rm g,s}\over \bar{B}_z^2/8\pi}\right)^{-1/2}. \label{eq:cminapp}
\end{equation}
Obviously, in practice such constraint disappears if $\beta_z^{\rm gt}/f_3 < 1$. The final expression is given in Equation (\ref{eq:cmin}).

\end{appendix}

\label{lastpage}
\end{document}